\documentclass[a4paper, 12pt]{article}
\usepackage[english]{babel}

\usepackage{paralist}

\usepackage[pdftex]{graphicx}

\usepackage[margin=15pt, font=small, labelfont=bf, tableposition=top, figureposition=bottom, skip=15pt]{caption}

\usepackage{amsmath}
\usepackage{amsthm}
\usepackage{amsfonts}
\usepackage{comment}
\usepackage{accents}

\usepackage[cp1250]{inputenc} 
\usepackage[IL2]{fontenc}

\usepackage{dcolumn}
\newcolumntype{d}{D{.}{.}{-1}}
\newcommand{\mc}[1]{\multicolumn{1}{c}{#1}}

\newcommand{\figscaleB}{0.54}

\newcommand{\bbe}{\mathbb{E}}
\newcommand{\bbp}{\mathbb{P}}

\newcommand{\bbq}{\mathbb{Q}}
\newcommand{\bbr}{\mathbb{R}}

\newcommand{\acal}{\mathcal{A}}

\newcommand{\dcal}{\mathcal{D}}

\newcommand{\fcal}{\mathcal{F}}
\newcommand{\gcal}{\mathcal{G}}

\newcommand{\ical}{\mathcal{I}}

\newtheorem{thm}{Theorem}

\newtheorem{de}{Definition}

\newtheorem{rem}{Remark}

\newcommand{\be}{\begin{eqnarray*}}
\newcommand{\ben}{\begin{eqnarray}}
\newcommand{\ee}{\end{eqnarray*}}
\newcommand{\een}{\end{eqnarray}}

\newcommand{\tendsto}{\operatornamewithlimits{\longrightarrow}}

\newcommand{\Tv}{T_{{\mathrm v}}}

\newcommand{\ubar}[1]{\underaccent{\bar}{#1}}

\begin{document}

\title{Valuation of Employee Stock Options (ESOs) by means
of Mean-Variance Hedging}

\author{Kamil Klad\'{i}vko \\
{\small \texttt{kladivko@gmail.com}} \\
%Department of Finance, Norwegian School of Economics \\ [4mm]
and \\ %[4mm]
Mihail Zervos\\
{\small \texttt{mihalis.zervos@gmail.com}} \\
Department of Mathematics, London School of Economics \\ [2mm]
}

\maketitle

\begin{abstract}
We consider the problem of ESO valuation in continuous
time.
In particular, we consider models that assume that
an appropriate random time serves as a proxy
for anything that causes the ESO's holder to exercise
the option early, namely, reflects the ESO holder's
job termination risk as well as early exercise
behaviour.
In this context, we study the problem of ESO valuation
by means of mean-variance hedging.
Our analysis is based on dynamic programming and uses
PDE techniques.
We also express the ESO's value that we derive as the
expected discounted payoff that the ESO yields with
respect to an equivalent martingale measure, which
does not coincide with the minimal martingale measure
or the variance-optimal measure.
Furthermore, we present a numerical study that illustrates
aspects or our theoretical results.
\bigskip

\bigskip

\end{abstract}

\newpage
\section{Introduction}

Employee stock options are call options granted
by a firm to its employees as a form of a benefit in addition
to salary.
Typical examples of ESO payoff functions $F$ include the
one of a call option with strike $K$, in which case,
$F(s) = (s - K)^+$,
as well as the payoff of a capped call option that pays out
no more than the double of its strike, in which case,
$F(s) = \bigl( s \wedge (2K) - K \bigr)^+$.
ESOs are typically long-dated options with maturities up
to several years.
Also, they typically have a {\em vesting period\/} of up
to several years, during which, they cannot be exercised.
After the expiry of their vesting period, they are of American
type.

The International Financial Reporting Standards Board
and Financial Accounting Standards Board require companies
to recognise an ESO as an expense in the income statement
at the moment the option is granted.
In particular, the fair-value principle is required for an ESO
valuation.
Such requirements as well as fundamental differences between
ESOs and standard traded options that we discuss below
have generated substantial interest in the development of
methodologies tailored to ESO valuation.

The risk-neutral valuation approach is a standard way of
pricing traded stock options.
However, this methodology does not apply to ESOs,
primarily, for the following two reasons:
\smallskip

({\em I\/})
If ESO holders have their jobs terminated (voluntarily or
because of being fired), they forfeit their unvested ESOs,
while they have a short time (typically, up to a few months)
to exercise their vested ESOs.
The possibility of  job termination presents an additional
uncertainty into the structure of an ESO, which is referred
to as the {\em job termination risk\/}.
\smallskip
 
({\em II\/})
ESOs are not allowed to be sold by their holders.
Furthermore, ESO holders face restrictions in trading their
employers' stocks.
Therefore, they cannot hedge the initial values of their
granted ESOs or use them as loss protections for
speculation on their underlying stock price declines.
These trading restrictions make ESO holders, who may be
in a need for liquidity or want to diversify their portfolios,
to exercise ESOs earlier than dictated by risk-neutrality.
The {\em early exercise behaviour\/} has been documented
in the empirical literature (e.g., see Huddart and
Lang~\cite{HL}), and has be explained theoretically by
means of expected utility maximisation techniques (e.g.,
see Sircar and Xiong~\cite{SX}).
\smallskip

A standard way of modelling the job termination risk and the
early exercise behaviour is by means of a Poisson process:
the ESO is {\em liquidated\/}, namely, exercised if vested and
in-the-money or forfeited if unvested or out-of-the money,
when the first jump of the Poisson process occurs.
This approach was introduced by Jennergren and
N\"{a}slund~\cite{JN}, and appears in the majority of ESO
valuation models, which can be classified in three main
groups.
Models in the first group use the first Poisson jump to capture
both the job termination risk and the voluntary decision of the
ESO holder to exercise (e.g., see
Rubinstein~\cite{R},
Carpenter~\cite{C}, 
Carr and Linetsky~\cite{CL}, and
Sircar and Xiong~\cite{SX}).
Models in the second group use the first Poisson jump to
capture the job termination risk, while they impose a barrier
that, when reached by the stock price, the ESO's exercise
is triggered  (e.g., see
Hull and White~\cite{HW}, and
Cvitanic, Wiener and Zapatero~\cite{CWZ}).
Models in these groups can be viewed as of an exogenous
or reduced form type.
Models in the third group, which can be viewed as of an
endogenous or structural type, use the first Poisson jump to
capture the job termination risk but determine the ESO
holder's early exercise strategy by maximising the holder's
utility of personal wealth (e.g., see
Leung and Sircar~\cite{LS}, and
Carpenter, Stanton and Wallace~\cite{CSW}).
The extensive use of a Poisson process in ESO valuation
has the attractive feature that, in its simplest form, it involves
a single parameter, namely, its intensity rate, which can
be estimated from historical data on ESO exercises
and forfeitures.
Indeed, the empirical study in Carpenter~\cite{C} shows that
a reduced form model with constant intensity of jumps can
perform as well or even better than more elaborate
structural models.

In this paper, we study a model that belongs to the first
family of models discussed above.
In particular, we consider an ESO with maturity $T$ that
is written on an underlying stock price process $S$, which
is modelled by a geometric Brownian motion.
We denote by $F$ the ESO's payoff function and we assume
that the ESO is vested at time $\Tv \in [0, T)$.
In the spirit of Carr and Linetsky~\cite{CL}, we model the
job termination risk as well as the voluntary decision of the
ESO holder to exercise, namely, the ESO's liquidation time,
by means of a random time $\eta$ with hazard rate that is
a function of the ESO's underlying stock price.
In this context, the ESO's valuation has to rely on an
incomplete market pricing methodology (see
Rheinlander and Sexton~\cite{RS11} for a textbook).

The {\em super-replication value\/} of the ESO is obtained
by viewing the random liquidation time $\eta$ as a
discretionary stopping time and treating the ESO as a
standard American option.
Accordingly, this value is given by
\ben
x^{\mathrm{sr}} = \sup _\varrho \, \bbe ^{\bbq_1} \left[ 
e^{-r (\varrho  \wedge T)} F(S_{\varrho  \wedge T}) \mathbf{1}
_{\{ \Tv \leq \varrho \}} \right] , \label{SRV}
\een
where $\bbq_1$ is the minimal martingale measure
and the supremum is taken over all stopping times $\varrho$
(see also Remark~\ref{MartMeasures}).
The super-replication value of an ESO is unrealistically
high because it does not take into account issues such
as the the job termination risk or the early exercise
behaviour discussed above.
Indeed, it is this observation that has given rise to the whole
research literature on the subject.

Another approach is to assign a value to the ESO by computing
its expected discounted payoff with respect to a martingale
measure.
For instance, we can assign the {\em risk-neutral value\/}
\ben
x^{\mathrm{rn}} = \bbe ^{\bbq_1} \left[ e^{-r (\eta \wedge T)}
F(S_{\eta  \wedge T}) \mathbf{1} _{\{ \Tv \leq \eta \}} \right]
\label{RNV}
\een
to the ESO, where $\bbq_1$ is the {\em minimal martingale
measure\/}, which, in the context we consider here, coincides
with the {\em variance-optimal martingale measure\/}
(see Remarks~\ref{MartMeasures} and~\ref{VO-MartMeas}).
Such a choice was proposed by Jennergren and
N\"{a}slund~\cite{JN} and Carr and Linetsky~\cite{CL}
by appealing to a {\em diversification\/} argument that
amounts to assuming that the jump risk is non-priced.
The reasoning behind this diversification assumption is that a
firm grants ESOs to a large number of employees, whose early
exercises and forfeitures are independent of each other.

Here, we study the mean-variance hedging of the
ESO's payoff.
The use of quadratic criteria to measure the quality of
a hedging strategy in continuous time has been proposed
by Bouleau and Lamberton~\cite{BL89}.
Mean-variance hedging was first studied in a specific
framework by Schweizer~\cite{Sc92} and has been
extensively studied since then by means of martingale
theory and $L^2$ projections
(e.g., see Pham~\cite{P00}, Schweizer~\cite{Sc01},
and \v{C}ern\'{y} and Kallsen~\cite{CK07}),
by means of PDEs
(e.g., see Bertsimas, Kogan and Lo~\cite{BKL})
as well as by means of BSDEs
(e.g., see Mania and Tevzadze~\cite{MT03}, and Jeanblanc,
Mania, Santacroce and Schweizer~\cite{JMSS12}).
Beyond such indicative references, we refer to
Schweizer~\cite{Sc10} for a survey of the vast literature
on the subject.
In particular, we consider the optimisation
problem
\ben
\text{minimise} \quad \bbe^\bbp \left[ \Bigl\{ e^{-r (\eta \wedge T)}
\bigl[ X_{\eta \wedge T}^{x,\pi} - F(S_{\eta \wedge T})
\mathbf{1} _{\{ \Tv \leq \eta \}} \bigr] \Bigr\}^2 \right] \quad
\text{over} \quad (x,\pi) , \label{P1}
\een
where $X^{x,\pi}$ is the value process of an admissible
self-financing portfolio strategy $\pi$ that starts with
initial endowment $x$ and $\bbp$ is the natural probability
measure.
We derive the solution to this problem by first solving the
problem
\ben
\text{minimise} \quad \bbe^\bbp \left[ \Bigl\{ e^{-r (\eta \wedge T)}
\bigl[ X_{\eta \wedge T}^{x,\pi} - F(S_{\eta \wedge T})
\mathbf{1} _{\{ \Tv \leq \eta \}} \bigr] \Bigr\}^2 \right] \quad
\text{over} \quad \pi , \label{P2}
\een
for any given initial endowment $x$ and then optimising
over $x$.
It turns out that the equivalent martingale measure for
the valuation of the ESO's payoff that arises from the
solution to the mean-variance optimisation problems
given by (\ref{P1})--(\ref{P2}) is different from the
coinciding in this context minimal martingale measure
and mean-variance martingale measure (see
Remarks~\ref{MartMeasures}--\ref{VO-MartMeas}).
This discrepancy can be attributed to the fact that market
incompleteness is due to the random time horizon
$\eta$.
Furthermore, it is worth noting that, although the solution to
(\ref{P2}) is time-consistent, the solution to (\ref{P1}) is
time-inconsistent (see also Remark~\ref{MV-ESO-value}).

The paper is organised as follows.
In Section~\ref{sec:CTE}, we formulate the problem
of ESO mean-variance hedging in continuous time.
In Section~\ref{sec:HJB}, we derive a {\em classical\/}
solution to the problem's Hamilton-Jacobi-Bellman
(HJB) equation, which takes the form of a nonlinear
parabolic partial differential equation (PDE).
We establish the main results on the mean-variance
hedging of an ESO's payoff in Section~\ref{sec:main}.
Finally, we present a numerical investigation in
Section~\ref{sec:numerics}.

%==========================================
\section{ESO mean-variance hedging}
\label{sec:CTE}

We build the model that we study in this section on a
complete probability space $(\Omega, \gcal, \bbp)$
carrying a standard one-dimensional Brownian motion
$W$ as well as an independent random variable $U$
that has the uniform distribution on $[0,1]$.
We denote by $(\fcal_t)$ the natural filtration of $W$,
augmented by the $\bbp$-negligible sets in $\gcal$.
In this probabilistic setting, we consider a firm whose stock price
process $S$ is modelled by the geometric Brownian motion
\ben
dS_t = \mu S_t \, dt + \sigma S_t \, dW_t , \quad S_0 = s
> 0 , \label{S}
\een
where $\mu$ and $\sigma \neq 0$ are given constants.
We assume that the firm can trade their own stock and
has access to a risk-free asset whose unit initialised price
is given by
\be
dB_t = r B_t \, dt , \quad B_0 = 1 ,
\ee
where $r \geq 0$ is a constant.
The value process of a self-financing portfolio with a position
in the firm's stock and a position in the risk-free asset that
starts with initial endowment $x$ has dynamics given by
\ben
dX_t = \bigl( rX_t + \sigma \vartheta \pi_t \bigr) \, dt +
\sigma \pi_t \, dW_t , \quad X_0 = x , \label{X}
\een
where $\pi_t$ is the amount of money invested in stock at time
$t$ and $\vartheta = (\mu - r) / \sigma$ is the market price
of risk.\footnote{Subject to suitable assumptions, the
analysis we develop can be trivially modified to allow
for $\mu$, $\sigma$ and $r$ to be functions of $t$ and
$S_t$. However, we opted against such a generalisation
because this would complicate substantially the notation.}
We restrict our attention to admissible portfolio strategies,
which are introduced by the following definition.

\begin{de} {\rm
Given a time horizon $T>0$, a portfolio process $\pi$ is
{\em admissible\/} if it is $(\fcal_t)$-progressively measurable
and
\ben
\bbe^\bbp \left[ \int _0^T \pi _t^2 \, dt \right] < \infty .
\label{pi-integr}
\een
We denote by $\acal_T$ the family of all such
portfolio processes.
} \mbox{}\hfill$\Box$
\end{de}

At time $0$, the firm issues an ESO that expires at time $T$
and is vested at time $\Tv \in [0, T)$, meaning that the ESO
can be exercised at any time between $\Tv$ and $T$.
We denote by $F(S)$ the payoff of the ESO.
The firm estimates that the holder of the ESO will either
exercise it or have their job terminated at a random
time $\eta$.
We model this time by 
\be
\eta = \inf \left\{ t \geq 0 \ \Big| \ \exp \left( - \int _0^t \ell
(u, S_u) \, du \right) \leq U \right\} ,
\ee
where the intensity function $\ell$ satisfies the
assumptions stated in Lemma~\ref{lem:fgh} below.
We note that the independence of $U$ and
$\fcal_\infty$ imply that
\be
\bbp (\eta > t \mid \fcal_t) = \bbp \left( U < \exp
\left( - \int _0^t \ell (u, S_u) \, du \right) \ \Big| \
\fcal_t \right) = \exp \left( - \int _0^t \ell (u, S_u)
\, du \right) .
\ee
We also denote by $(\gcal_t)$ the filtration derived by
rendering right-continuous the filtration defined by
$\fcal_t \vee \sigma \bigl( \{ \eta \leq s \} , \
s \leq t \bigr)$, for $t \geq 0$.
It is a standard exercise of the credit risk theory to
show that the process $M$ defined by
\ben
M_t = {\bf 1} _{\{ \eta \leq t \}} - \int _0^{t \wedge \eta}
\ell (u, S_u) \, du \label{M}
\een
is a $(\gcal_t)$-martingale.

\begin{rem} \label{MartMeasures} {\rm
We will consider probability measures that are
equivalent to $\bbp$ and are parametrised by
$(\fcal_t)$-predictable processes $\gamma > 0$
satisfying suitable integrability conditions (see
Blanchet-Scalliet, El~Karoui and Martellini~\cite{BSEKM},
and Blanchet-Scalliet and Jeanblanc~\cite{BSJ04}).
Given such a process, the solution to the SDE
\be
dL_t^\gamma = (\gamma_{t-} - 1) L_{t-}^\gamma \, dM_t
- \vartheta L_t^\gamma \, dW_t ,
\ee
where $M$ is the $(\gcal_t)$-martingale defined by
(\ref{M}), which is given by
\be
L_t^\gamma = \exp \left( {\bf 1} _{\{ \eta \leq t \}} \ln
\gamma_\eta - \int _0^{t \wedge \eta} \ell (u, S_u)
\bigl( \gamma_u -1 \bigr) \, du - \frac{1}{2} \vartheta^2 t
- \vartheta W_t \right) ,
\ee
defines an exponential martingale.
If we denote by $\bbq_\gamma$ the probability measure
on $(\Omega, \gcal_T)$ that has Radon-Nikodym
derivative with respect to $\bbp$ given by
$\left. \frac{d\bbq_\gamma}{d\bbp} \right| _{\gcal_T}
= L_T^\gamma$, then Girsanov's theorem implies that
the process $\bigl( \tilde{W}_t , \ t \in [0,T] \bigr)$ is a
standard Brownian motion under $\bbq_\gamma$,
while the process $\bigl( \tilde{M}_t , \ t \in [0,T] \bigr)$
is a martingale under $\bbq_\gamma$, where
\be
\tilde{W}_t = \vartheta t + W_t \quad \text{and}
\quad \tilde{M}_t = {\bf 1} _{\{ \eta \leq t \}} -
\int _0^{t \wedge \eta} \ell (u, S_u) \gamma_u
\, du , \quad \text{for } t \in [0,T] .
\ee
Furthermore, the dynamics of the stock price process
are given by
\be
dS_t = r S_t \, dt + \sigma S_t \, d\tilde{W}_t ,
\quad \text{for } t \in [0,T] , \quad S_0 = s
> 0 ,
\ee
while the conditional distribution of $\eta$ is
given by
\be
\bbq_\gamma (\eta > t \mid \fcal_t) = \exp \left(
- \int _0^t \ell (u, S_u) \gamma_u \, du
\right) , \quad \text{for } t \in [0,T] .
\ee
In this context, the choice $\gamma = 1$ gives rise to
the minimal martingale measure, which coincides with
the variance-optimal martingale measure
(see Blanchet-Scalliet, El Karoui and
Martellini~\cite{BSEKM}, and Szimayer~\cite{Sz}).
} \mbox{}\hfill$\Box$ \end{rem}

In the probabilistic setting that we have developed,
we consider a firm whose objective is to invest an initial amount
$x$ in a self-financing portfolio with a view to hedging against
the payoff $F(S_\eta)$ that they have to pay the ESO's holder
at time $\eta$.
To this end, a risk-neutral valuation approach is not possible
due to the market's incompleteness.
We therefore consider minimising the {\em expected squared
hedging error\/}, which gives rise to the following stochastic
control problem.

Given an ESO with expiry date $T$ that is vested
at time $\Tv \in [0,T)$,
the objective is to minimise the performance criterion
\begin{align}
J_{T,x,s} (\pi) & = \bbe^\bbp \left[ \bigl\{ e^{-r\eta}
X_\eta \bigr\} ^2 \mathbf{1} _{\{ 0 \leq \eta < \Tv \}}
+ \Bigl\{ e^{-r (\eta \wedge T)} \bigl[ X_{\eta \wedge T}
- F(S_{\eta \wedge T}) \bigr] \Bigr\}^2 \mathbf{1}
_{\{ \Tv \leq \eta \}} \right] \nonumber \\
& = \bbe^\bbp \left[ e^{-2r (\eta \wedge T)} \bigl[ X_{\eta \wedge T}
- F(S_{\eta \wedge T}) {\mathbf 1} _{\{ \Tv \leq \eta \}} \bigr]
^2 \right] \label{eq:objCT0}  
\end{align}
over all admissible self-financing portfolio strategies.
In view of the underlying probabilistic setting, this
performance index admits the expression
\ben
J_{T,x,s} (\pi) = \bbe^\bbp \left[ \int_0^T e^{-\Lambda_t}
\ell (t, S_t) \bigl[ X_t - F(S_t) \mathbf{1} _{\{ \Tv \leq t \}}
\bigr]^2 \, dt + e^{-\Lambda_T} \bigl[ X_T - F(S_T) \bigr]^2
\right] . \label{eq:objCT}
\een
where
\be
\Lambda_t = 2rt + \int _0^t \ell (u, S_u) \, du .
\ee
The value function of the resulting optimisation problem
is defined by
\ben
v(T,x,s) = \inf _{\pi \in \acal_T} J_{T,x,s} (\pi) .
\label{v}
\een

%==========================================
\section{A classical solution to the HJB equation}
\label{sec:HJB}

In view of standard stochastic control theory, the
value function $v$ should identify with a solution $w$
to the HJB PDE 
\begin{align}
-w_\tau (\tau,x,s) + \inf _\pi \left\{ \frac{1}{2} \sigma^2 \pi^2
w_{xx} (\tau,x,s) + \sigma^2 s \pi w_{xs} (\tau,x,s) + (rx +
\sigma \vartheta \pi) w_x (\tau,x,s) \right\} & \nonumber \\
\mbox{} + \frac{1}{2} \sigma^2 s^2 w_{ss} (\tau,x,s) + \mu
s w_s (\tau,x,s) - \bigl( 2r + \lambda (\tau,s) \bigr) w(\tau,x,s)
& \nonumber \\
\mbox{} + \lambda (\tau,s) \bigl[ x - F(s) \mathbf{1}
_{\{ \tau \leq T - \Tv \}} \bigr]^2 & = 0 \label{HJB}
\end{align}
that satisfies the initial condition
\ben
w(0,x,s) = \bigl[ x - F(s) \bigr]^2 , \label{HJB-BC}
\een
where the independent variable $\tau = T-t$ denotes
time-to-maturity and
\ben
\lambda (\tau,s) = \ell (T-\tau, s) . \label{ell-lambda}
\een
If the function $w(\tau, \cdot, s)$ is convex for all
$(\tau,s) \in [0,T] \times \bbr_+$, then the infimum in this
PDE is achieved by
\ben
\pi^\dagger (\tau,x,s) = - \frac{\sigma s w_{xs} (\tau,x,s)
+ \vartheta w_x (\tau,x,s)} {\sigma w_{xx} (\tau,x,s)} .
\label{pi-dagger}
\een
and (\ref{HJB}) is equivalent to
\begin{align}
-w_\tau (\tau,x,s) - \frac{\bigl( \sigma s w_{xs} (\tau,x,s)
+ \vartheta w_x (\tau,x,s) \bigr)^2}{2 w_{xx} (\tau,x,s)} +
\frac{1}{2} \sigma^2 s^2 w_{ss} (\tau,x,s) + rx w_x (\tau,x,s)
& \nonumber \\
\mbox{} + \mu s w_s (\tau,x,s) - \bigl( 2r + \lambda (\tau,s)
\bigr) w(\tau,x,s)  + \lambda (\tau,s) \bigl[ x - F(s) \mathbf{1}
_{\{ \tau \leq T - \Tv \}}
\bigr]^2 & = 0 . \label{HJB1}
\end{align}
In view of the quadratic structure of the problem we consider,
we look for a solution to this PDE of the form
\ben
w(\tau,x,s) = f(\tau,s) \bigl[ x - g(\tau,s) \bigr]^2 +
h(\tau,s) , \label{w-sol}
\een
for some functions $f$, $g$ and $h$.
Substituting this expression for $w$ in (\ref{HJB1}), we can
see that the functions $f$, $g$ and $h$ should satisfy
the PDEs
\begin{align}
-f_\tau (\tau,s) + \frac{1}{2} \sigma^2 s^2 f_{ss} (\tau,s)
+ \mu s f_s (\tau,s) - \lambda (\tau,s) f (\tau,s) + \lambda
(\tau,s) \hspace{22mm} \nonumber \\
\mbox{} - \frac{\bigl( \sigma s f_s (\tau,s) + \vartheta f(\tau,s)
\bigr)^2}{f (\tau,s)} & = 0 , \label{f} \\
-g_\tau (\tau,s) + \frac{1}{2} \sigma^2 s^2 g_{ss} (\tau,s)
+ rs g_s (\tau,s) - \left( r + \frac{\lambda (\tau,s)}{f(\tau,s)}
\right) g (\tau,s) \hspace{25mm} \nonumber \\
\mbox{} + \frac{\lambda (\tau,s) F(s) \mathbf{1}
_{\{ \tau \leq T - \Tv \}} }{f(\tau,s)} & = 0 , \label{g} \\
-h_\tau (\tau,s) + \frac{1}{2} \sigma^2 s^2 h_{ss}
(\tau,s) + \mu s h_s (\tau,s) - \bigl( 2r + \lambda (\tau,s)
\bigr) h (\tau,s) \hspace{25mm} & \nonumber \\
\mbox{}  + \lambda (\tau,s) \bigl[ F(s) \mathbf{1}
_{\{ \tau \leq T - \Tv \}} - g(\tau,s) \bigr]^2 & = 0
\label{h}
\end{align}
in $(0,T] \times (0, \infty)$, with initial conditions
\ben
f(0,s) = 1 , \quad g(0,s) = F(s) \quad \text{and} \quad
h(0,s) = 0 . \label{fgh-BC}
\een

The following result addresses the solvability
of these PDEs as well as certain estimates we will
need.

\begin{thm} \label{lem:fgh}
Suppose that the functions $\lambda$ and $F$
are $C^1$ and there exist constants $\bar{\lambda},
K_F > 0$ and $\xi \geq 1$ such that
\begin{gather}
0 \leq \lambda (\tau,s) + s \bigl| \lambda_s (\tau,s)
\bigr| \leq \bar{\lambda} \quad \text{for all } \tau, s
> 0 \label{lambda-bounds} \\
\text{and} \quad
0 \leq F(s) + s \bigl| F'(s) \bigr| \leq K_F \left(
1 + s^\xi \right) \quad \text{for all } s > 0 .
\label{F-bounds}
\end{gather}
The following statements hold true:
\smallskip

\noindent{\rm (I)}
The PDE (\ref{f}) with the corresponding boundary
condition in (\ref{fgh-BC}) has a $C^{1,2}$ solution
such that
\begin{gather}
\ubar{K}_f \leq f(\tau, s) \leq \bar{K}_f \quad \text{for all }
\tau \in [0,T] \text{ and } s > 0 \label{f-bounds} \\
\text{and} \quad \bigl| f_s (\tau,s) \bigr| \leq
\bar{K}_f s^{-1} \quad \text{for all } \tau \in
[0,T] \text{ and } s > 0 , \label{f_s-bounds}
\end{gather}
for some constants $\ubar{K}_f = \ubar{K}_f (T) > 0$
and $\bar{K}_f = \bar{K}_f (T) > \ubar{K}_f$.

\noindent{\rm (II)}
The PDE (\ref{g}) with the corresponding boundary
condition in (\ref{fgh-BC}) has a $C^{1,2}$ solution
such that
\begin{gather}
0 \leq g(\tau, s) \leq K_g \left( 1 + s^\xi \right) \quad
\text{for all } \tau \in [0,T] \text{ and } s > 0
\label{g-bounds} \\
\text{and} \quad
\bigl| g_s (\tau,s) \bigr| \leq K_g \left( 1 + s^\xi \right)
s^{-1} \quad \text{for all } \tau \in [0,T] \text{ and }
s > 0 , \label{g_s-bounds}
\end{gather}
for some constant $K_g = K_g (T) > 0$.
\smallskip

\noindent{\rm (III)}
The PDE (\ref{h}) with the corresponding boundary
condition in (\ref{fgh-BC}) has a $C^{1,2}$ solution
such that
\ben
0 \leq h(\tau, s) \leq K_h \left( 1 + s^{2\xi} \right)
\quad \text{for all } \tau \in [0,T] \text{ and } s > 0
, \label{h-bounds}
\een
for some constant $K_h = K_h (T) > 0$.
\end{thm}
\noindent
{\bf Proof.}
%=======
We establish each of the parts sequentially. 
\smallskip

\underline{\em Proof of (I).}
If we define
\ben
f(\tau,s) = \phi^{-1} (\tau,s) , \quad \text{for } \tau
\in [0, T] \text{ and } s > 0 , \label{f-phi}
\een
then we can see that $f$ satisfies the PDE (\ref{f}) in
$(0,T] \times (0, \infty)$ with the corresponding initial
condition in (\ref{fgh-BC}) if and only if $\phi$ satisfies
the PDE
\ben
- \phi_\tau (\tau, s) + \frac{1}{2} \sigma^2 s^2 \phi_{ss}
(\tau,s) + (r - \sigma \vartheta) s \phi_s (\tau, s) - \bigl(
\lambda (\tau,s) \phi (\tau, s) - \lambda (\tau,s) - \vartheta^2
\bigr) \phi (\tau,s) = 0 \label{phi0-PDE}
\een
in $(0,T] \times (0, \infty)$ with initial condition
\ben
\phi (0,s) = 1 , \quad \text{for } s>0 . \label{phi0-BC}
\een
Furthermore, if we write
\ben
\phi (\tau, s) = e^{(\bar{\lambda} + \vartheta^2) \tau}
\varphi (\tau, \ln s) , \quad \text{for } \tau \in [0, T]
\text{ and } s > 0 , \label{phi-varphi}
\een
for some function $(\tau,z) \mapsto \varphi (\tau,z)$,
where $\bar{\lambda}$ is as in (\ref{lambda-bounds}),
then we can check that $\phi$ satisfies the PDE
(\ref{phi0-PDE}) with initial condition (\ref{phi0-BC})
if and only if $\varphi$ satisfies the PDE
\begin{align}
- \varphi_\tau (\tau,z) + \frac{1}{2} \sigma^2 \varphi_{zz}
(\tau,z) + \left( r - \sigma \vartheta - \frac{1}{2} \sigma^2
\right) \varphi_z (\tau,z) & \nonumber \\
- \left( e^{(\bar{\lambda} + \vartheta^2) \tau} \lambda
(\tau,e^z) \varphi (\tau,z) + \bar{\lambda} - \lambda
(\tau,e^z) \right) \varphi (\tau,z) & = 0 \label{phi-PDE}
\end{align}
in $(0,T] \times \bbr$ with initial condition
\ben
\varphi (0,z) = 1 , \quad \text{for } z \in \bbr .
\label{phi-BC}
\een
To solve this nonlinear PDE, we consider the family of
linear PDEs
\ben
- \varphi_\tau^\psi (\tau,z) + \frac{1}{2} \sigma^2
\varphi_{zz}^\psi (\tau,z) + \left( r - \sigma \vartheta -
\frac{1}{2} \sigma^2 \right) \varphi_z^\psi (\tau,z) -
\delta^\psi (\tau,z) \varphi^\psi (\tau,z) = 0
\label{phi-PDE-par}
\een
in $(0,T] \times (0, \infty)$ with initial condition
\ben
\varphi^\psi (0,z) = 1 , \quad \text{for } z \in \bbr ,
\label{phi-BC-par}
\een
which is parametrised by smooth positive functions $\psi$,
where
\ben
\delta^\psi (\tau,z) = e^{(\bar{\lambda} + \vartheta^2) \tau}
\lambda (\tau,e^z) \psi (\tau,z) + \bar{\lambda} - \lambda
(\tau,e^z) \geq 0 , \quad \text{for } \tau \in [0,T] \text{ and }
z \in \bbr . \label{delta}
\een
In particular, we note that a solution to (\ref{phi-PDE})
satisfies (\ref{phi-PDE-par}) for $\psi = \varphi$.

Consider a $C^{1,2}$ function $\psi$ satisfying
\ben
0 \leq \psi (\tau,z) \leq 1 \quad \text{and} \quad
|\psi_z (\tau,z)| \leq C_1 \quad \text{for all } \tau \in [0,T]
\text{ and } z \in \bbr , \label{psi-ass}
\een
for some constant $C_1 = C_1 (T)$.
The properties of such a function and the assumptions
on $\lambda$ imply that there exists a unique $C^{1,2}$
function $\varphi^\psi$ of polynomial growth that solves
the Cauchy problem (\ref{phi-PDE-par})--(\ref{phi-BC-par})
(see Friedman~\cite[Section~6.4]{F2} or
Friedman~\cite[Section~1.7]{F1}).
In view of the Feynman-Kac formula (see
Friedman~\cite[Section~6.5]{F2} or Karatzas
and Shreve~\cite[Theorem~5.7.6]{KS}), this function
admits the probabilistic representation
\ben
\varphi^\psi (\tau,z) = \bbe \left[ \exp \left( - \int _0^\tau
\delta^\psi (\tau-u, Z_u) \, du \right) \Big| \ Z_0 = z \right]
\in (0,1] , \quad \text{for } \tau \in [0,T] \text{ and }
z \in \bbr , \label{FK-phi-a}
\een
where $Z$ is the Brownian motion with drift given by
\ben
dZ_t = \left( r - \sigma \vartheta - \frac{1}{2} \sigma^2
\right) dt + \sigma \, dB_t , \label{Z}
\een
for some standard one-dimensional Brownian motion
$B$.
The assumptions on $\psi$ imply that $\varphi _z^\psi$
is $C^{1,2}$ (see Friedman~\cite[Section~3.5]{F1}).
Differentiating (\ref{phi-PDE-par}), we can see that
$\varphi _z^\psi$ satisfies
\begin{align}
- \varphi_{\tau z}^\psi (\tau,z) + \frac{1}{2} \sigma^2
\varphi_{zzz}^\psi (\tau,z) + \left( r - \sigma \vartheta -
\frac{1}{2} \sigma^2 \right) \varphi_{zz}^\psi (\tau,z)
- \delta^\psi (\tau,z) \varphi_z^\psi (\tau,z) & \nonumber \\
- \left( e^{(\bar{\lambda} + \vartheta^2) \tau}
e^z \lambda_s (\tau, e^z) \psi (\tau,z) + e^{(\bar{\lambda}
+ \vartheta^2) \tau} \lambda (\tau, e^z) \psi_z (\tau,z) -
e^z \lambda_s (\tau, e^z) \right) \varphi^\psi (\tau,z)
& = 0 \nonumber
\end{align}
in $(0,T] \times \bbr$.
Using the Feynman-Kac formula, Jensen's inequality,
(\ref{lambda-bounds}), (\ref{psi-ass}) and (\ref{FK-phi-a}),
we can see that
\begin{align}
\bigl| \varphi_z^\psi (\tau,z) \bigr|
& \leq \bbe \biggl[ \int_0^\tau \exp \left( - \int _0^u
\delta^\psi (\tau-q, Z_q) \, dq \right) \nonumber \\
& \qquad\qquad\quad \times
\Bigl( e^{(\bar{\lambda} + \vartheta^2) (\tau-u)}
e^{Z_u} \bigl| \lambda_s (\tau-u, e^{Z_u}) \bigr| \psi
(\tau-u,Z_u) \nonumber \\
& \qquad\qquad\quad \qquad \mbox{}
+ e^{(\bar{\lambda} + \vartheta^2) (\tau-u)}
\lambda (\tau-u, e^{Z_u}) \bigl| \psi_z (\tau-u,Z_u) \bigr|
\nonumber \\
& \qquad\qquad\quad \qquad \mbox{}
+ e^{Z_u} \bigl| \lambda_s (\tau-u, e^{Z_u}) \bigr| \Bigr)
\varphi^\psi (\tau-u,{Z_u}) \, du \ \Big| \ Z_0 = z \biggr]
\nonumber \\
& \leq \int_0^\tau \bar{\lambda} \left( (1 + C_1)
e^{(\bar{\lambda} + \vartheta^2) (\tau-u)} + 1
\right) du \nonumber \\
& \leq \frac{\bar{\lambda} (1 + C_1)}{\bar{\lambda}
+ \vartheta^2} e^{(\bar{\lambda} + \vartheta^2) T}
+ \bar{\lambda} T \quad \text{for all } \tau \in [0,T]
\text{ and } z \in \bbr , \nonumber
\end{align}
where $Z$ is the Brownian motion with drift given by
(\ref{Z}).
It follows that $\varphi^\psi$ inherits all of the
properties that we have assumed for $\psi$ above.
Furthermore, Schauder's interior estimates for parabolic
PDEs implies that, given any bounded open interval
$\ical \subset \bbr$,
\ben
\| \varphi^\psi \| _{1+a,2+a}^{(0,T) \times \ical}
\mbox{} \leq C_2 \sup _{\tau \in (0,T) , \, z \in \ical} \bigl|
\varphi^\psi (\tau,z) \bigr| \leq C_2 , \label{S-est}
\een
where
\begin{gather}
\| \varphi \| _{1+a,2+a}^\dcal \mbox{} =
\mbox{} \| \varphi \| _a^\dcal \mbox{} +
\mbox{} \| \varphi_t \| _a^\dcal \mbox{} +
\mbox{} \| \varphi_z \| _a^\dcal \mbox{} +
\mbox{} \| \varphi_{zz} \| _a^\dcal , \nonumber \\
\mbox{} \| \varphi \| _a^\dcal \mbox{} =
\sup _{(t,z) \in \dcal} |\varphi (t,z)| +
\sup _{\substack{(t,z), (t',z') \in \dcal \\ (t,z) \neq (t',z')}}
\frac{|\varphi (t,z) - \varphi (t',z')|}{|t-t'|^{a/2} +
|z-z'|^a} , \nonumber
\end{gather}
and $C_2$ depends only on $a$ and $\ical$
(see Friedman~\cite[Section~3.2]{F1}).

To proceed further, we denote by $\varphi^{(0)}$
the solution to (\ref{phi-PDE-par})--(\ref{phi-BC-par})
for $\psi \equiv 0$ and by $\varphi^{(j+1)}$
the solution to (\ref{phi-PDE-par})--(\ref{phi-BC-par})
for $\psi = \varphi^{(j)}$ and $j \geq 0$.
By appealing to a simple induction argument,
we can see that each $\varphi^{(j)}$ has all
of the properties that we assumed for $\psi$ in
the previous paragraph.
Therefore, all of the functions $\varphi^{(j)}$,
$j \geq 0$, satisfy the estimates (\ref{S-est}) for
the same constant $C_2$.
This observation and the Arzel\`{a}-Ascoli
theorem imply that there exist a $C^{1,2}$ function
$\varphi$ and a sequence of natural numbers
$(j_n)$ such that
\be
\varphi^{(j_n)} \tendsto_{n \rightarrow \infty} \varphi ,
\quad
\varphi_t^{(j_n)} \tendsto_{n \rightarrow \infty} \varphi_t
, \quad
\varphi_z^{(j_n)} \tendsto_{n \rightarrow \infty} \varphi_z
\quad \text{and} \quad
\varphi_{zz}^{(j_n)} \tendsto_{n \rightarrow \infty}
\varphi_{zz} ,
\ee
uniformly on compacts.
Such a limiting function is a solution to
(\ref{phi-PDE-par})--(\ref{phi-BC-par}) for $\psi = \varphi$,
namely, a solution to the nonlinear PDE (\ref{phi-PDE})
that satisfies the initial condition (\ref{phi-BC}).
Furthermore, $\varphi _z$ satisfies the PDE
\begin{align}
- \varphi_{\tau z} (\tau,z) + \frac{1}{2} \sigma^2
\varphi_{zzz} (\tau,z) + \left( r - \sigma \vartheta -
\frac{1}{2} \sigma^2 \right) \varphi_{zz} (\tau,z) &
\nonumber \\
- \left( 2 e^{(\bar{\lambda} + \vartheta^2) \tau}
\lambda (\tau, e^z) \varphi (\tau,z) + \bar{\lambda} -
\lambda (\tau,e^z) \right) \varphi_z (\tau,z) &
\nonumber \\
- e^z \lambda_s (\tau, e^z) \Bigl( e^{(\bar{\lambda}
+ \vartheta^2) \tau} \varphi (\tau,z) - 1 \Bigr)
\varphi (\tau,z) & = 0 \nonumber
\end{align}
in $(0,\infty) \times \bbr$.
It follows that the function $\phi$ given by (\ref{phi-varphi})
is such that $\phi_s$ satisfies the PDE
\begin{align}
- \phi_{\tau s} (\tau,s) + \frac{1}{2} \sigma^2 s^2
\phi_{sss} (\tau,s) + \left( r - \sigma \vartheta +
\sigma^2 \right) s \phi_{ss} (\tau,s) & \nonumber \\
- \left( 2 \lambda (\tau, s) \phi (\tau,s) - \lambda (\tau, s)
- r - \vartheta^2 + \sigma \vartheta \right) \phi_s
(\tau,s) - \lambda_s (\tau, s) \left( \phi (\tau,s) - 1 \right)
\phi (\tau,s) & = 0 \label{phi_s}
\end{align}
in $(0,T] \times (0,\infty)$, as well as the boundary
condition
\ben
\phi_s (0,s) = 0 , \quad \text{for } s>0 . \label{phi_s-BC}
\een

To establish (\ref{f-bounds})--(\ref{f_s-bounds}), we first
note that (\ref{FK-phi-a}) yields the representations
\begin{align}
\varphi^{(0)} (\tau,z) & = \bbe \left[ \exp \left( - \int _0^\tau
\delta^0 (\tau-u, Z_u) \, du \right) \Big| \ Z_0 = z \right]
\nonumber \\
\text{and} \quad
\varphi^{(j+1)} (\tau,z) & = \bbe \left[ \exp \left( - \int _0^\tau
\delta^{\varphi^{(j)}} (\tau-u, Z_u) \, du \right) \Big| \
Z_0 = z \right] , \label{FK-phi-j}
\end{align}
for $\tau \in [0,T]$, $z \in \bbr$ and $j \geq 0$,
where $Z$ is the Brownian motion with drift given by
(\ref{Z}).
Combining these expressions with the definition (\ref{delta})
of the functions $\delta^{\varphi^{(j)}}$, we can see that
\begin{align}
& \varphi^{(0)} > \varphi^{(1)} \quad \Rightarrow \quad
- \delta^{\varphi^{(0)}} < - \delta^{\varphi^{(1)}}
\quad \Rightarrow \quad
\varphi^{(1)} < \varphi^{(2)} , \nonumber \\
& \varphi^{(1)} < \varphi^{(2)} \quad \Rightarrow \quad
- \delta^{\varphi^{(1)}} > - \delta^{\varphi^{(2)}}
\quad \Rightarrow \quad
\varphi^{(2)} > \varphi^{(3)} , \nonumber \\
& \varphi^{(0)} > \varphi^{(2)} \quad \Rightarrow \quad
- \delta^{\varphi^{(0)}} < - \delta^{\varphi^{(2)}}
\quad \Rightarrow \quad
\varphi^{(1)} < \varphi^{(3)} , \nonumber \\
\text{and} \qquad
& \varphi^{(1)} < \varphi^{(3)} \quad \Rightarrow \quad
- \delta^{\varphi^{(1)}} > - \delta^{\varphi^{(3)}}
\quad \Rightarrow \quad
\varphi^{(2)} > \varphi^{(4)} . \nonumber
\end{align}
Iterating these observations, we can see that
the sequence of functions $(\varphi^{(2j)})$
is strictly decreasing, while the sequence
of functions $(\varphi^{(2j+1)})$ is strictly increasing.
It follows that
\ben
\varphi^{(1)} \leq \varphi \leq \varphi^{(0)} \leq
1 . \label{bars-phi-ineqs}
\een
In view of (\ref{lambda-bounds}) and (\ref{FK-phi-j}),
we calculate
\begin{align}
\varphi^{(1)} (\tau,z) & \geq \bbe \left[ \exp \left( - \int _0^\tau
\left( e^{(\bar{\lambda} + \vartheta^2) (\tau-u)} \lambda
(\tau-u, e^{Z_u}) + \bar{\lambda} - \lambda (\tau-u, e^{Z_u})
\right) du \right) \ \Big| \ Z_0 = z \right] \nonumber \\
& \geq \exp \left( - \bar{\lambda} \int _0^\tau
e^{(\bar{\lambda} + \vartheta^2) (\tau-u)} \, du \right)
\nonumber \\
& \geq \exp \left( - e^{(\bar{\lambda} + \vartheta^2) \tau}
\right) \qquad \text{for all } \tau \in [0,T] \text{ and } z \in
\bbr . \label{varphi_2-est}
\end{align}
Combining the inequalities (\ref{bars-phi-ineqs}) and
(\ref{varphi_2-est}) with (\ref{f-phi}) and (\ref{phi-varphi}),
we obtain (\ref{f-bounds}).

Using the Feynman-Kac formula, Jensen's inequality,
(\ref{lambda-bounds}), (\ref{phi-varphi}) and
(\ref{bars-phi-ineqs}), we can see that the solution to
(\ref{phi_s})--(\ref{phi_s-BC}) satisfies
\begin{align}
\bigl| \phi_s (\tau,s) \bigr|
& \leq \bbe \biggl[ \int_0^\tau \exp \left( - \int _0^u
\left( 2 \lambda (\tau-q, \bar{S}_q) \phi (\tau-q,s) - \lambda
(\tau-q, \bar{S}_q) - r - \vartheta^2 + \sigma \vartheta \right)
dq \right) \nonumber \\
& \qquad \qquad\quad \times
\bigl| \lambda_s (\tau-u, \bar{S}_u) \bigr| \left( \phi
(\tau-u,\bar{S}_u) + 1 \right) \phi (\tau-u,\bar{S}_u)
\, du \ \Big| \ \bar{S}_0 = s \biggr] \nonumber \\
& \leq \bar{\lambda} e^{(\bar{\lambda} + \vartheta^2) \tau}
\, \bbe \left[ \int_0^\tau e^{(r-\sigma \vartheta) u}
\bar{S}_u^{-1} \left( e^{(\bar{\lambda} + \vartheta^2)
(\tau-u)} + 1 \right) du \ \Big| \ \bar{S}_0 = s \right]
\nonumber \\
& = \bar{\lambda} e^{(\bar{\lambda} + \vartheta^2) \tau}
s^{-1} \int_0^\tau \left( e^{(\bar{\lambda} + \vartheta^2)
(\tau-u)} + 1 \right) du \nonumber \\
& \leq 2 e^{2(\bar{\lambda} + \vartheta^2) \tau}
s^{-1} \qquad \text{for all } \tau \in [0,T] \text{ and }
s > 0 , \label{phi_s-bound}
\end{align}
where $\bar{S}$ is the geometric Brownian
motion given by
\be
d\bar{S}_t = (r - \sigma \vartheta + \sigma^2)
\bar{S}_t \, dt + \sigma \bar{S}_t \, dB_t ,
\ee
for a standard one-dimensional Brownian motion
$B$.
Combining this estimate with (\ref{f-phi}),
(\ref{phi-varphi}), (\ref{bars-phi-ineqs}) and
(\ref{varphi_2-est}), we obtain
\begin{align}
\bigl| f_s (\tau,s) \bigr| & = \frac{\bigl| \phi_s
(\tau,s) \bigr|}{\phi^2 (\tau,s)} \leq 2 \exp \left(
2 e^{(\bar{\lambda} + \vartheta^2) \tau}
\right) s^{-1} \quad \text{for all } \tau \in [0,T]
\text{ and } s > 0 , \nonumber
\end{align}
and (\ref{f_s-bounds}) follows.
\smallskip

\underline{\em Proof of (II).}
If we write
\be
g (\tau, s) = \tilde{g} (\tau, \ln s) , \quad \text{for } \tau \geq
0 \text{ and } s > 0 ,
\ee
for some function $(\tau,z) \mapsto \tilde{g} (\tau,z)$, then
$g$ satisfies the PDE (\ref{g}) in $(0,T] \times (0, \infty)$
with the corresponding initial condition in (\ref{fgh-BC})
if and only if $\tilde{g}$ satisfies the PDE
\begin{align}
- \tilde{g}_\tau (\tau,z) + \frac{1}{2} \sigma^2
\tilde{g}_{zz} (\tau,z) + \left( r - \frac{1}{2} \sigma^2
\right) \tilde{g}_z (\tau,z) & \nonumber \\
\quad - \left( r + e^{(\bar{\lambda} + \vartheta^2) \tau}
\lambda (\tau, e^z) \varphi (\tau,z) \right) \tilde{g} (\tau,z)
+ e^{(\bar{\lambda} + \vartheta^2) \tau} \lambda (\tau, e^z)
\varphi (\tau,z) F(e^z) \mathbf{1} _{\{ \tau \leq T - \Tv \}}
& = 0 , \nonumber
\end{align}
in $(0,T] \times \bbr$, where $\varphi$ is introduced
by (\ref{phi-varphi}), with initial condition
\be
\tilde{g} (0,z) = F(e^z) , \quad \text{for } z \in \bbr .
\ee
In view of the assumptions on $\lambda$, $F$, and
the smoothness and boundedness of $\varphi$ that
we have established above, there exists a unique
$C^{1,2}$ function $\tilde{g}$ of polynomial growth
that solves this Cauchy problem (see
Friedman~\cite[Section~1.7]{F1}).

In view of the Feynman-Kac formula
(Friedman~\cite[Section~6.5]{F2} or Karatzas
and Shreve~\cite[Theorem~5.7.6]{KS}), the solution
to the PDE (\ref{g}) with the corresponding initial
condition in (\ref{fgh-BC}) admits the probabilistic
expression
\begin{align}
g(\tau,s) = \bbe \biggl[ & \int _0^\tau \exp \left( -
\int _0^u \left( r + \frac{\lambda (\tau-q, \tilde{S}_q)}
{f(\tau-q, \tilde{S}_q)} \right) dq \right)
\frac{\lambda (\tau-u, \tilde{S}_u) F(\tilde{S}_u)
\mathbf{1} _{\{ \tau-u \leq T - \Tv \}}}{f(\tau-u,\tilde{S}_u)}
\, du \nonumber \\
& + \exp \left( - \int _0^\tau \left( r +
\frac{\lambda (\tau-u, \tilde{S}_u)}{f(\tau-u, \tilde{S}_u)}
\right) du \right) F(\tilde{S}_\tau) \ \Big| \ \tilde{S}_0
= s \biggr] \nonumber \\
\geq \mbox{} & 0 \qquad \text{for all } \tau \in [0,T]
\text{ and } s > 0, \label{g-prob}
\end{align}
where $\tilde{S}$ is the geometric Brownian
motion given by
\ben
d\tilde{S}_t = r \tilde{S}_t \, dt + \sigma
\tilde{S}_t \, dB_t , \label{g-prob-S}
\een
for a standard one-dimensional Brownian motion
$B$.
Using (\ref{lambda-bounds})--(\ref{f-bounds}),
we can see that this expression implies that
\begin{align}
g(\tau,s) & \leq \bbe \left[ \int _0^\tau \bar{\lambda}
\ubar{K}_f^{-1} K_F (1 + \tilde{S}_u^\xi) \, du + K_F
(1 + \tilde{S}_\tau^\xi) \ \Big| \ S_0 = s \right] \nonumber \\
& = \int _0^\tau \bar{\lambda} \ubar{K}_f^{-1} K_F \left(
1 + s^\xi e^{\left( \frac{1}{2} \sigma^2 \xi (\xi - 1) +
r \xi \right) u} \right) du \nonumber \\
& \quad + K_F \left( 1 + s^\xi e^{\left( \frac{1}{2}
\sigma^2 \xi (\xi - 1) + r \xi \right) \tau} \right)
\qquad \text{for all } \tau \in [0,T] \text{ and }
s > 0 . \nonumber
\end{align}
It follows that $g$ admits an upper bound as in
(\ref{g-bounds}).

Similarly to the proof of~(I) above, we can verify
that $g_s$ satisfies the PDE
\ben
-g_{\tau s} (\tau,s) + \frac{1}{2} \sigma^2 s^2 g_{sss}
(\tau,s) + (r + \sigma^2) s g_{ss} (\tau,s) - \lambda (\tau, s)
\phi (\tau,s) g_s (\tau,s) + \gamma (\tau,s) & = 0 ,
\label{g_s-PDE}
\een
in $(0,T] \times \bbr$, where $\phi$ is as in
(\ref{f-phi}) and
\begin{align}
\gamma (\tau,s) = \mbox{} & - \bigl[ \lambda_s (\tau, s)
\phi (\tau,s) + \lambda (\tau, s) \phi_s (\tau,s) \bigr]
\bigl[ g(\tau,s) - F(s) \mathbf{1} _{\{ \tau \leq T - \Tv \}} \bigr]
\nonumber \\
& + \lambda (\tau, s) \phi (\tau,s) F'(s) \mathbf{1}
_{\{ \tau \leq T - \Tv \}} . \nonumber
\end{align}
Using the relevant bounds in (\ref{lambda-bounds}),
(\ref{F-bounds}), (\ref{g-bounds}), (\ref{bars-phi-ineqs})
and (\ref{phi_s-bound}), we calculate
\begin{align}
\bigl| \gamma (\tau,s) \bigr| \leq \mbox{} & \left(
\bar{\lambda} e^{(\bar{\lambda} + \vartheta^2) \tau}
s^{-1} + 2\bar{\lambda} e^{3(\bar{\lambda} +
\vartheta^2) \tau} s^{-1} \right) (K_g + K_F) (1 + s^\xi)
+ \bar{\lambda} e^{(\bar{\lambda} + \vartheta^2) \tau}
K_F (1 + s^\xi) s^{-1} \nonumber \\
\leq \mbox{} & K_\gamma \left( 1 + s^\xi \right) s^{-1}
\qquad \text{for all } \tau \in [0,T] \text{ and }
s > 0 , \nonumber
\end{align}
for some constant $K_\gamma = K_\gamma (T)
> 0$.
If we denote by $\hat{S}$ the geometric Brownian
motion given by
\be
d\hat{S}_t = (r + \sigma^2) \hat{S}_t \, dt + \sigma
\hat{S}_t \, dB_t ,
\ee
where $B$ is a standard one-dimensional Brownian
motion, then (\ref{g_s-PDE}), the Feynman-Kac
formula, Jensen's inequality and the estimate
for $\gamma$ derived above yield
\begin{align}
\bigl| g_s (\tau,s) \bigr| & \leq \bbe \left[ \int _0^\tau
\exp \left( - \int _0^u \lambda (\tau-q, \hat{S}_q)
\phi (\tau-q, \hat{S}_q) \, dq \right) \bigl| \gamma
(\tau-u,\hat{S}_u) \bigr| \, du \ \Big| \ \hat{S}_0 = s
\right] \nonumber \\
& \leq \bbe \left[ \int _0^\tau K_\gamma \left( \hat{S}_u^{-1}
+ \hat{S}_u^{\xi -1} \right) \ \Big| \ \hat{S}_0 = s \right]
\nonumber \\
& = K_\gamma \int _0^\tau \left( s^{-1} e^{-ru} +
s^{\xi - 1} e^{\left( \frac{1}{2} \sigma^2 (\xi - 2) +
(r+\sigma^2)) (\xi -1) \right) u} \right) du \quad
\text{for all } \tau \in [0,T] \text{ and } s > 0 .\nonumber
\end{align}
It follows that $|g_s|$ admits a bound as in
(\ref{g_s-bounds}).
\smallskip

\underline{\em Proof of (III).}
We can show that the PDE (\ref{h}) in $(0,T] \times (0, \infty)$
with the corresponding initial condition in (\ref{fgh-BC}) has
a $C^{1,2}$ solution in the same way as in the proof of~(II).
Using the Feynman-Kac formula once again, we can
see that this solution admits the probabilistic expression
\begin{align}
h(\tau,s) & = \bbe \biggl[ \int _0^\tau \exp \left( -
\int _0^u \bigl( 2r + \lambda (\tau-q, S_q) \bigr) \, dq
\right) \nonumber \\
\mbox{} & \qquad \qquad \ \times
\lambda (\tau-u, S_u) \bigl[ F(S_u) \mathbf{1}
_{\{ \tau-u \leq T - \Tv \}} - g(\tau-u,S_u) \bigr]^2
\, du \biggr] \geq 0 , \label{h-prob}
\end{align}
where $S$ is the geometric Brownian motion given
by (\ref{S}).
In view of (\ref{lambda-bounds}) and (\ref{g-bounds}),
we can see that this expression implies that
\begin{align}
h(\tau,s) & \leq \bbe \left[ \int _0^\tau 2 \lambda
(\tau-u, S_u) \bigl[ F^2(S_u) + g^2 (\tau-u,S_u) \bigr]
\, du \ \Big| \ S_0 = s \right] \nonumber \\
& \leq \bbe \left[ \int _0^\tau 4 \bar{\lambda} \left[
K_F^2 (1 + S_u^{2\xi}) + K_g^2 (1 + S_u^{2\xi})
\right] du \ \Big| \ S_0 = s \right]
\nonumber \\
& = \int _0^\tau 4 \bar{\lambda} \biggl[ K_F^2 \left( 1 
+ s^{2\xi} e^{\left( \sigma^2 \xi (2\xi - 1) + 2\mu \xi
\right) u} \right) + K_g^2 \left( 1 + s^{2\xi} e^{\left(
\sigma^2 \xi (2\xi - 1) + 2\mu \xi \right) u} \right)
\biggr] du \nonumber
\end{align}
for all $\tau \in [0,T]$ and $s>0$, and the claim in
(\ref{h-bounds}) follows.
\mbox{}\hfill$\Box$

%==========================================
\section{The main results on ESO mean-variance
hedging}
\label{sec:main}

In view of (\ref{pi-dagger}) and (\ref{w-sol}), we can
see that the optimal portfolio strategy is given by
\begin{align}
\pi_t^\star = \pi^\dagger (T-t,X_t^\star,S_t)
& = - \alpha_t \bigl[ X_t^\star - g(T-t,S_t) \bigr] +
S_t g_s (T-t,S_t) = - \alpha_t X_t^\star + \beta_t ,
\label{pi-star}
\end{align}
where $\pi^\dagger$ is the function defined by
(\ref{pi-dagger}),
\ben
\alpha_t = \frac{\sigma S_t f_s (T-t,S_t) + \vartheta
f(T-t,S_t)}{\sigma f(T-t,S_t)} , \qquad
\beta_t = \alpha_t g(T-t,S_t) + S_t g_s (T-t,S_t) ,
\label{alpha}
\een
and $X^\star$ is the associated solution to (\ref{X}).

The following result presents the solution to the
control problem associated with ESO mean-variance
hedging.

\begin{thm} \label{prop:VT}
Consider the stochastic control problem defined by
(\ref{S}), (\ref{X}), (\ref{eq:objCT}) and (\ref{v}),
and suppose that the assumptions of
Lemma~\ref{lem:fgh} hold true.
The problem's value function $v$ identifies with
the solution $w$ to the HJB PDE (\ref{HJB})--(\ref{HJB-BC})
that is as in (\ref{w-sol})--(\ref{fgh-BC}), namely,
\ben
v(T,x,s) = w(T,x,s) \quad \text{for all } x \in \bbr \text{ and }
s > 0 . \label{v=w}
\een
Furthermore, the portfolio strategy $\pi^\star$ defined
by (\ref{pi-star})--(\ref{alpha}) is optimal.
\end{thm}
\noindent
{\bf Proof.}
%=======
\noindent
We fix any initial condition $(x,s)$ and any admissible portfolio
$\pi \in \acal_T$.
Using It\^{o}'s formula, we calculate
\begin{align}
\int_0^t e^{-\Lambda_u} \ell (u,S_u) \bigl[ X_u - F(S_u)
\mathbf{1} _{\{ \Tv \leq u \}} \bigr]^2 \, du & \nonumber \\
\mbox{} + e^{-\Lambda_t} w(T-t, X_t , S_t) & = w(T,x,s)
+ A_t +M_t , \quad \text{for } t \in [0,T] ,
\end{align}
where
\begin{align}
A_t & = \int _0^{t \wedge T} e^{-\Lambda_u} \biggl[
- w_\tau (T-u, X_u, S_u) + \frac{1}{2} \sigma^2 \pi_u^2
w_{xx} (T-u, X_u, S_u) \nonumber \\
& \qquad\qquad\qquad\
+ \sigma^2 S_u \pi_u w_{xs} (T-u, X_u, S_u) +
\frac{1}{2} \sigma^2 S_u^2 w_{ss} (T-u, X_u, S_u)
\nonumber \\
& \qquad\qquad\qquad\
+ (rX_u + \sigma \vartheta \pi_u) w_x (T-u, X_u, S_u)
+ \mu S_u w_s (T-u, X_u, S_u) \nonumber \\
& \qquad\qquad\qquad\
- \bigl( 2r + \ell (u,S_u) \bigr) w(T-u, X_u, S_u) + \ell
(u,S_u) \bigl[ X_u - F(S_u) \mathbf{1} _{\{ \Tv \leq u \}}
\bigr]^2 \biggr] \, du \nonumber
\end{align}
and
\be
M_t = \sigma \int_0^{t \wedge T} e^{-\Lambda_u} \bigl[
\pi_u w_x (T-u, X_u, S_u) + S_u w_s (T-u, X_u, S_u)
\bigr] \, dW_u .
\ee
Since $w$ satisfies the PDE (\ref{HJB}) and the
boundary condition (\ref{HJB-BC}), we can see that this
identity implies that
\begin{align}
\bbe^\bbp \biggl[ & \int_0^{T \wedge T_n} e^{-\Lambda_u}
\ell (u,S_u) \bigl[ X_u - F(S_u) \mathbf{1} _{\{ \Tv \leq u \}}
\bigr]^2 \, du \nonumber \\
& + e^{-\Lambda_T} \bigl[ X_T - F(S_T) \bigr]^2
{\bf 1} _{\{ T \leq T_n \}} + e^{-\Lambda_{T_n}} w(T - T_n,
X_{T_n}, S_{T_n}) {\bf 1} _{\{ T_n < T \}} \biggr] \geq
w(T,x,s) , \label{VT-1}
\end{align}
where $(T_n)$ is any sequence of localising stopping
times for the local martingale $M$.

In view of the estimates in (\ref{f-bounds}),
(\ref{g-bounds}) and (\ref{h-bounds}), we can see
that
\begin{align}
\bigl| w(\tau,x,s) \bigr| & \leq 2 f(\tau,s) \bigl[ x^2
+ g^2(\tau,s) \bigr] + h(\tau,s) \nonumber \\
& \leq K_w \left( 1 + x^2 + s^{2\xi} \right)
\quad \text{for all } \tau \in [0,T] , \, x \in \bbr
\text{ and } s > 0 , \nonumber
\end{align}
for some constants $K_w = K_w (T) > 0$ and
$\xi_w > 0$.
On the other hand, the admissibility condition
(\ref{pi-integr}), Fubini's theorem, Jensen's inequality
and It\^{o}'s isometry imply that
\begin{align}
\bbe^\bbp \bigl[ X _t^2 \bigr] & = \bbe^\bbp \left[
e^{2rt} \left( x + \sigma \vartheta \int _0^t e^{-ru}
\pi_u \, du + \sigma \int _0^t e^{-ru} \pi_u \, dW_u
\right) ^2 \right] \nonumber \\
& \leq 9 e^{2rt} \left( x^2 + \sigma^2 \vartheta^2
\, \bbe^\bbp \left[ \left( \int _0^t e^{-ru} \pi_u \, du
\right)^2 \right] + \sigma^2 \, \bbe^\bbp \left[ \left(
\int _0^t e^{-ru} \pi_u \, dW_u \right)^2 \right] \right)
\nonumber \\
& \leq 9 e^{2rt} \left( x^2 + \sigma^2 (\vartheta^2 t
+1) \, \bbe^\bbp \left[ \int _0^t e^{-2ru} \pi_u^2 \, du
\right] \right) \nonumber \\
& < \infty \nonumber
\end{align}
These results imply that the random variable
$\sup _{t \in [0,T]} \bigl| w(T-t, X_t, S_t) \bigr|$ is
integrable.
We can therefore pass to the limit as $n \rightarrow \infty$
in (\ref{VT-1}) using the monotone and the dominated
convergence theorems to obtain
\begin{align}
J_{T,x,s} (\pi) & \equiv \bbe^\bbp \left[ \int_0^T e^{-\Lambda_t}
\ell (t,S_t) \bigl[ X_t - F(S_t) \mathbf{1} _{\{ \Tv \leq t \}}
\bigr]^2 \, dt + e^{-\Lambda_T} \bigl[ X_T - F(S_T) \bigr]^2
\right] \nonumber \\
& \geq w(T,x,s) . \label{VT-2}
\end{align}
Since the initial condition $(x,s)$ and the portfolio strategy
$\pi \in \acal_T$ have been arbitrary, it follows that
\ben
v(T,x,s) \geq w(T,x,s) \quad \text{for all } T>0 , \, x \in \bbr
\text{ and } s > 0 . \label{VT-3}
\een

To prove the reverse inequality and establish (\ref{v=w})
as well as the optimality of the portfolio strategy
$\pi^\star$ defined by (\ref{pi-star})--(\ref{alpha}), we first
show that $\pi^\star$ is admissible, namely, $\pi^\star
\in \acal_T$.
To this end, we first note that
\begin{align}
\bbe^\bbp \left[ {X_t^\star}^2 \right] \leq 25 \biggl( & x^2 +
\bbe^\bbp \left[ \left( \int _0^t (r - \sigma \vartheta \alpha_u)
X_u^\star \, du \right) ^2 \right] + \sigma^2 \vartheta^2 \,
\bbe^\bbp \left[ \left( \int _0^t \beta_u \, du \right) ^2 \right]
\nonumber \\
& + \sigma^2 \, \bbe^\bbp \left[ \int _0^t \alpha_u^2
{X_u^\star}^2 \, du \right] + \sigma^2 \, \bbe^\bbp \left[
\int _0^t \beta_u^2 \, du \right] \biggr) , \label{pistar-adm}
\end{align}
where we have also used It\^{o}'s isometry.
The estimates (\ref{f-bounds})--(\ref{f_s-bounds}) imply
that there exists a constant $K_\alpha = K_\alpha (T)$
such that
\ben
|\alpha_t| \leq \frac{\sigma S_t \bigl| f_s (T-t,S_t) \bigr|
+ \vartheta f(T-t,S_t)}{\sigma f(T-t,S_t)} \leq K_\alpha
\quad \text{for all } t \in [0,T] , \label{alpha-B}
\een
while the estimates (\ref{g-bounds})--(\ref{g_s-bounds})
imply that there exists a constant $K_\beta = K_\beta (T)$
such that
\ben
\beta _t^2 \leq 2\alpha_t^2 g^2(T-t,S_t) + 2S_t^2 g_s^2
(T-t,S_t) \leq K_\beta (1 + S_t^{2\xi}) \quad \text{for all }
t \in [0,T] . \label{beta-B}
\een
Using these inequalities, we can see that, e.g.,
\begin{align}
\bbe^\bbp \left[ \int _0^t \alpha_u^2 {X_u^\star}^2
\, du \right] & \leq K_\alpha^2 \int _0^t \bbe^\bbp \left[
{X_u^\star}^2 \right] du \qquad \text{for all } t \in [0,T]
\nonumber \\
\text{and} \quad
\bbe^\bbp \left[ \left( \int _0^t \beta_u \, du \right) ^2 \right]
& \leq T \, \bbe^\bbp \left[ \int _0^t \beta_u^2 \, du \right]
\nonumber \\
& \leq K_\beta T \left( 1 + \int _0^t \bbe^\bbp \left[
S_u^{2\xi} \right] \, du \right)
\leq C_1 (1 + s^{2\xi}) \quad \text{for all } t \in [0,T] ,
\nonumber
\end{align}
where $C_1 = C_1 (T)$ is a constant.
In view of these inequalities and similar ones for
the other terms, we can see that (\ref{pistar-adm})
implies that there exists $C_2 = C_2 (T,s)$ such that
\be
\bbe^\bbp \left[ {X_t^\star}^2 \right] \leq C_2 + C_2
\int _0^t \bbe^\bbp \left[ {X_u^\star}^2 \right] du .
\ee
It follows that
\be
\bbe^\bbp \left[ {X_t^\star}^2 \right] \leq C_2 e^{C_2 t}
\quad \text{for all } t \in [0,T] ,
\ee
thanks to Gr\"{o}nwall's inequality.
Combining this result with (\ref{alpha-B}) and
(\ref{beta-B}), we obtain
\begin{align}
\bbe^\bbp \left[ \int _0^T {\pi _t^\star}^2 \, dt \right]
& \leq 2 \, \bbe^\bbp \left[ \int _0^T \left( \alpha_t^2
{X _t^\star}^2 + \beta_t^2 \right) dt \right]
\nonumber \\
& \leq 2 \int _0^T \left( K_\alpha^2 C_2 e^{C_2 t}
+ K_\beta \left( 1 +  \bbe^\bbp \left[ S_t^{2\xi} \right]
\right) \right) dt < \infty , \nonumber
\end{align}
and the admissibility of $\pi^\star$ follows.

Finally, it is straightforward to check that the portfolio
strategy $\pi^\star$ defined by (\ref{pi-star})--(\ref{alpha})
is such that (\ref{VT-1}) as well as (\ref{VT-2}) hold
true with equality, which combined with the inequality
(\ref{VT-3}), implies that $\pi^\star$ is optimal
and that (\ref{v=w}) holds true.
\mbox{}\hfill$\Box$

\begin{rem} \label{MV-ESO-value} {\rm
Given an ESO such as the one we have considered,
the self-financing portfolio's initial endowment $X_0^\star
= x^\star$ that minimises the expected squared hedging
error is equal to $g(T,S_0)$, which is the ESO's {\em
mean-variance hedging value at time\/} 0.
It is worth noting that the optimal portfolio strategy
$\pi^\star$ that starts with initial endowment $X_0^\star
= g(T,S_0)$ has value process $X^\star$ such that
$X_t^\star \neq g(T-t,S_t)$ (this can be seen by a
comparison of the dynamics of the processes $X^\star$
and $\bigl( g(T-t, S_t) , \ t \in [0,T] \bigr)$).
We can therefore view $g(T-t, S_t)$ as the ESO's
{\em mark-to-market mean-variance hedging value at time}
$t$.
} \mbox{}\hfill$\Box$ \end{rem}

\begin{rem} \label{MartMeas} {\rm
We can express the ESO's value $g(T,s)$ at time
0 as its expected with respect to a martingale measure
discounted cost to the firm.
To this end, we consider the exponential martingale
$(L_t , \ t \in [0,T])$ that solves the SDE
\be
dL_t = \bigl( f^{-1} (T-t,S_t) - 1 \bigr) L_{t-} \, dM_t
- \vartheta L_t \, dW_t ,
\ee
where $M$ is the $(\gcal_t)$-martingale defined by
(\ref{M}), and is given by
\begin{align}
L_t = \exp \biggl( & - {\bf 1} _{\{ \eta \leq t \}} \ln
f(T - \eta, S_\eta) \nonumber \\
& - \int _0^{t \wedge \eta} \ell (u, S_u) \bigl(
f^{-1} (T-u,S_u) -1 \bigr) \, du - \frac{1}{2} \vartheta^2
t - \vartheta W_t \biggr) , \quad \text{for } t \in
[0,T] . \nonumber
\end{align}
If we denote by $\bbq$ the probability measure on
$(\Omega, \gcal_T)$ that has Radon-Nikodym
derivative with respect to $\bbp$ given by
$\left. \frac{d\bbq}{d\bbp} \right| _{\gcal_T} =
L_T$, then Girsanov's theorem implies that
the process $\bigl( \tilde{W}_t , \ t \in [0,T] \bigr)$
is a standard Brownian motion under $\bbq$, while
the process $\bigl( \tilde{M}_t , \ t \in [0,T] \bigr)$
is a martingale under $\bbq$, where
\be
\tilde{W}_t = \vartheta t + W_t \quad \text{and}
\quad \tilde{M}_t = {\bf 1} _{\{ \eta \leq t \}} -
\int _0^{t \wedge \eta} \ell (u, S_u) \bigl(
f^{-1} (T-u,S_u) -1 \bigr) \, du , \quad \text{for }
t \in [0,T] .
\ee
Furthermore, the dynamics of the stock price process
are given by
\be
dS_t = r S_t \, dt + \sigma S_t \, d\tilde{W}_t ,
\quad \text{for } t \in [0,T] , \quad S_0 = s
> 0 ,
\ee
while the conditional distribution of $\eta$ is
given by
\be
\bbq (\eta > t \mid \fcal_t) = \exp \left( - \int _0^t
\frac{\ell (u, S_u)}{f(T-u,S_u)} \, du \right) ,
\quad \text{for } t \in [0,T] .
\ee
In view of these observations and the Feynman-Kac
formula (see also (\ref{g-prob})--(\ref{g-prob-S}) in
Appendix~I), we can see that
\begin{align}
g(T,s) & = \bbe^\bbq \biggl[ \int _0^T \exp \left( -
\int _0^t \left( r + \frac{\ell (u, S_u)}{f(T-u,S_u)}
\right) du \right) \frac{\ell (t, S_t) F(S_t)
\mathbf{1} _{\{ \Tv \leq t \}}}{f(T-t,S_t)} \, dt
\nonumber \\
& \qquad \qquad
+ \exp \left( - \int _0^T \left( r +
\frac{\ell (u, S_u)}{f(T-u,S_u)} \right) du \right)
F(S_T) \biggr] \nonumber \\
& = \bbe^\bbq \Bigl[ e^{-r (\eta \wedge T)}
F(S_{\eta \wedge T}) \mathbf{1} _{\{ \Tv \leq \eta \}}
\Bigr] , \label{g-probQ}
\end{align}
as claimed at the beginning of the remark.
} \mbox{}\hfill$\Box$ \end{rem}

\begin{rem} \label{VO-MartMeas} {\rm
Suppose that $\sigma \vartheta = \mu -r = 0$.
In this special case, we can check that the constant function
$f \equiv 1$ satisfies the PDE (\ref{f}), the PDE
(\ref{g}) takes the form
\ben
-g_\tau (\tau,s) + \frac{1}{2} \sigma^2 s^2 g_{ss} (\tau,s)
+ rs g_s (\tau,s) - \bigl( r + \lambda (\tau,s) \bigr) g (\tau,s)
+ \lambda (\tau, s) F(s) \mathbf{1} _{\{ \tau \leq T - T_v \}}
= 0 , \label{g-RN}
\een
the mean-variance hedging value of the ESO is given by
\ben
g(T,s) = \bbe^\bbp \left[ \int _0^T e^{- \int_0^t (r + \ell
(u, S_u)) \, du} \ell (t, S_t) F(S_t) {\mathbf 1}
_{\{ \Tv \leq t \}} \, dt + e^{- \int_0^T (r + \ell
(u, S_u)) \, du} F(S_T) \right] , \label{ESO-value-RN}
\een
and the optimal portfolio is given by
\be
\pi_t^\star = S_t g_s (T-t,S_t) .
\ee
Jennergren and N\"{a}slund~\cite{JN} and
Carr and Linetsky~\cite{CL} assume that the ESO's
exercise risk can be {\em diversified away\/} and
{\em propose\/} (\ref{ESO-value-RN}) to be the value
of the ESO.
Furthermore, they derive the PDE (\ref{g-RN})
with boundary condition $g(0,s) = F(s)$ by
appealing to the Feynman-Kac theorem, and they
solve it for the special cases that arise when
$\Tv = 0$,
\be
\ell (T-\tau, s) \equiv \lambda (\tau, s) = \lambda_f
+ \lambda_e {\bf 1} _{\{ s > K \}} \quad \text{or} \quad
\ell (T-\tau, s) \equiv \lambda (\tau, s) = \lambda_f
+ \lambda_e (\ln s - \ln K)^+
\ee
and $F(s) = (s-K)^+$, for some constants
$\lambda_f, \lambda_e, K >0$.
Effectively, this approach amounts to choosing the
minimal martingale measure for the valuation of the
ESO (see also Remark~\ref{MartMeasures}).
Indeed, in the general case, i.e., when $\mu \neq r$,
the function $g$ given by (\ref{ESO-value-RN}) for
$\mu = r$ identifies with the function $g$ given by
\begin{align}
g(T,s) & = \bbe^{\bbq_1} \left[ \int _0^T
e^{- \int_0^t (r + \ell (u, S_u)) \, du}
\ell (t, S_t) F(S_t) {\mathbf 1} _{\{ \Tv \leq t \}}
\, dt + e^{- \int_0^T (r + \ell (u, S_u)) \, du}
F(S_T) \right] \nonumber \\
& = \bbe^{\bbq_1} \Bigl[ e^{-r (\eta \wedge T)}
F(S_{\eta \wedge T}) {\mathbf 1} _{\{ \Tv \leq \eta \}}
\Bigr] , \nonumber
\end{align}
where $\bbq_1$ is the probability measure with
Radon-Nikodym derivative with respect to $\bbp$
given by $\left. \frac{d\bbq_1}{d\bbp} \right|
_{\gcal_T} = L_T^1 \equiv \exp \left( - \frac{1}{2}
\vartheta^2 T - \vartheta W_T \right)$.
} \mbox{}\hfill$\Box$ \end{rem}

\begin{rem} \label{rem:infinite} {\rm
({\em The infinite time horizon case.\/})
In many cases, ESOs are very long-dated.
It is therefore of interest to consider the form
that the solution to the problem we have studied
takes as $T \rightarrow \infty$.
In this case, if $\Tv = 0$ and $\ell$ does not
depend explicitly on time, then the solution to the
control problem becomes stationary, namely, it does
not depend on time.
In particular, the value function $v_\infty$ identifies
with the function $w_\infty$ defined by
\be
w_\infty (x,s) = f_\infty (s) \bigl[ x - g_\infty (s)
\bigr]^2 + h_\infty (s) ,
\ee
and the optimal portfolio strategy is given by
\be
\pi_t^\star = - \frac{\sigma S_t f_\infty' (S_t) +
\vartheta f_\infty (S_t)}{\sigma f_\infty (S_t)}
\bigl[ X_t^\star - g_\infty (S_t) \bigr] + S_t
g_\infty' (S_t) ,
\ee
where $X^\star$ is the associated solution to
(\ref{X}), and the functions $f_\infty$, $g_\infty$,
$h_\infty$ are appropriate solutions to the ODEs
\begin{align}
\frac{1}{2} \sigma^2 s^2 f_\infty'' (s) + \mu s f_\infty' (s)
- \lambda (s) f_\infty (s) + \lambda (s) -
\frac{\bigl( \sigma s f_\infty' (s) + \vartheta f_\infty (s)
\bigr)^2}{f_\infty (s)} & = 0 , \label{foo} \\
\frac{1}{2} \sigma^2 s^2 g_\infty'' (s)
+ rs g_\infty' (s) - \left( r + \frac{\lambda (s)}{f_\infty (s)}
\right) g_\infty (s) + \frac{\lambda (s) F(s)}{f_\infty (s)}
& = 0 , \label{goo} \\
\frac{1}{2} \sigma^2 s^2 h_\infty''
(s) + \mu s h_\infty' (s) - \bigl( 2r + \lambda (s)
\bigr) h_\infty (s) + \lambda (s) \bigl[ F(s) -
g_\infty (s) \bigr]^2 & = 0 . \label{hoo}
\end{align}
The nonlinear ODE (\ref{foo}) with general $\lambda$
requires a separate analysis that goes beyond the
scope of this article.
For the purposes of this remark, we therefore assume
that
\begin{gather}
\ell \equiv \lambda > 0 \text{ is a constant} , \quad
0 \leq F(s) \leq K_F \left( 1 + s^\xi \right) \text{ for all }
s > 0 , \label{F-ineq-oo} \\
\lambda + \vartheta^2 > \left( r + \frac{1}{2} \sigma^2 \xi
\right) (\xi - 1) \quad \text{and} \quad \lambda >
2r(\xi - 1) + 2\sigma \vartheta \xi + \sigma^2 \xi (2\xi - 1) ,
\label{par-ineq-oo}
\end{gather}
where $K_F > 0$ and $\xi \geq 1$ are constants.
Note that, if $\xi = 1$, then the inequalities (\ref{par-ineq-oo})
are equivalent to the simpler
\ben
\lambda > 2 \left( \mu - r + \frac{1}{2} \sigma^2 \right) .
\label{par-ineq-oo-1}
\een
For constant $\lambda$,  we can verify that the solution to
(\ref{f}) that satisfies the corresponding boundary
condition in (\ref{fgh-BC}) is given by
\ben
f(\tau,s) = \frac{\lambda}{\lambda + \vartheta^2}
+ \frac{\vartheta^2}{\lambda + \vartheta^2}
e^{- (\lambda + \vartheta^2) \tau} . \label{f-lam-const}
\een
In view of this observation, we can see that the constant
function given by $f_\infty (s) = \frac{\lambda}
{\lambda + \vartheta^2}$, for $s>0$, trivially satisfies
(\ref{foo}).
Furthermore, (\ref{g-probQ}) and (\ref{h-prob}) suggest
that the functions given by
\begin{align}
g_\infty (s) & = (\lambda + \vartheta^2) \, \bbe^\bbq
\left[ \int _0^\infty e^{- (r +\lambda + \vartheta^2) t}
F(S_t) \, dt \right] \nonumber \\
\text{and} \quad
h_\infty (s) & = \lambda \, \bbe^\bbp \left[ \int _0^\infty
e^{-(2r + \lambda) t} \bigl[ F(S_t) - g_\infty (S_t)
\bigr]^2 \, dt \right] \nonumber
\end{align}
should satisfy the ODEs (\ref{goo}) and (\ref{hoo}).
Note that the conditions in (\ref{F-ineq-oo}) and
(\ref{par-ineq-oo}) are sufficient for these functions to be
real-valued because
\begin{align}
g_\infty (s) & \leq (\lambda + \vartheta^2) K_F
\int _0^\infty e^{- (r + \lambda + \vartheta^2) t} \left(
1 + \bbe^\bbq \bigl[ S_t^\xi \bigr] \right) dt \nonumber \\
& = (\lambda + \vartheta^2) K_F \int _0^\infty
e^{- (r + \lambda + \vartheta^2) t} \left( 1 + s^\xi
e^{\left( \frac{1}{2} \sigma^2 \xi (\xi - 1)
+ r \xi \right) t} \right) dt \nonumber \\
& \leq K_{g_\infty} (1 + s^{\xi}) , \nonumber
\end{align}
where $K_{g_\infty}$ is a constant, and
\begin{align}
h_\infty (s) & \leq 4 \lambda \left( K_F^2 + K_{g_\infty}^2
\right) \int _0^\infty e^{-(2r + \lambda) t} \left( 1 + \bbe^\bbp
\bigr[ S_t^{2\xi} \bigr] \right) dt \nonumber \\
& = 4 \lambda \left( K_F^2 + K_{g_\infty}^2 \right) \int
_0^\infty e^{-(2r + \lambda) t} \left( 1 + s^{2\xi}
e^{\left( \sigma^2 \xi (2\xi - 1) + 2(\sigma \vartheta + r)
\xi \right) t} \right) dt \nonumber \\
& < \infty . \nonumber
\end{align}
In view of standard analytic expressions of resolvents
(e.g., see Knudsen, Meister and Zervos
\cite[Proposition~4.1]{KMZ} or Lamberton and
Zervos~\cite[Theorem 4.2]{LZ}), these functions admit
the analytic expressions
\begin{align}
g_\infty (s) & = \frac{\lambda + \vartheta^2} {\sigma^2
(n_g - m_g)} \left[ s^{m_g} \int _0^s u^{- m_g - 1} F(u) \,
du + s^{n_g} \int _s^\infty u^{- n_g - 1} F(u) \,
du \right] \nonumber \\
\text{and} \quad
h_\infty (s) & = \frac{\lambda} {\sigma^2
(n_h - m_h)} \biggl[ s^{m_h} \int _0^s
u^{- m_h - 1} \bigl[ F(u) - g_\infty (u)
\bigr]^2 \, du \nonumber \\
& \qquad \qquad \qquad \qquad
+ s^{n_h} \int _s^\infty u^{- n_h - 1} \bigl[
F(u) - g_\infty (u) \bigr]^2 \, du \biggr] ,
\nonumber
\end{align}
where the constants $m_g < 0 < n_g$ and
$m_h < 0 < n_h$ are defined by
\begin{align}
m_g , n_g & = - \frac{r - \frac{1}{2} \sigma^2}{\sigma^2}
\mp \sqrt{\left( \frac{r - \frac{1}{2} \sigma^2}{\sigma^2}
\right)^2 + \frac{2}{\sigma^2} \left( r + \frac{\lambda^2}
{\lambda + \vartheta^2} \right)} , \nonumber \\
m_h , n_h & = - \frac{\mu - \frac{1}{2} \sigma^2}{\sigma^2}
\mp \sqrt{\left( \frac{\mu - \frac{1}{2} \sigma^2}{\sigma^2}
\right)^2 + \frac{2 (2r + \lambda)}{\sigma^2}} . \nonumber
\end{align}
We can check that these functions indeed satisfy the
ODEs (\ref{goo}) and (\ref{hoo}) by direct substitution.
It is worth noting that we can use these expressions
to calculate $g_\infty$ and $h_\infty$ in closed
analytic form for a most wide range of choices
for $F$.
} \mbox{}\hfill$\Box$ \end{rem}

%==========================================
\section{Numerical investigation}
\label{sec:numerics}

\begin{figure}[!b]
\begin{center}
\scalebox{\figscaleB}{\includegraphics{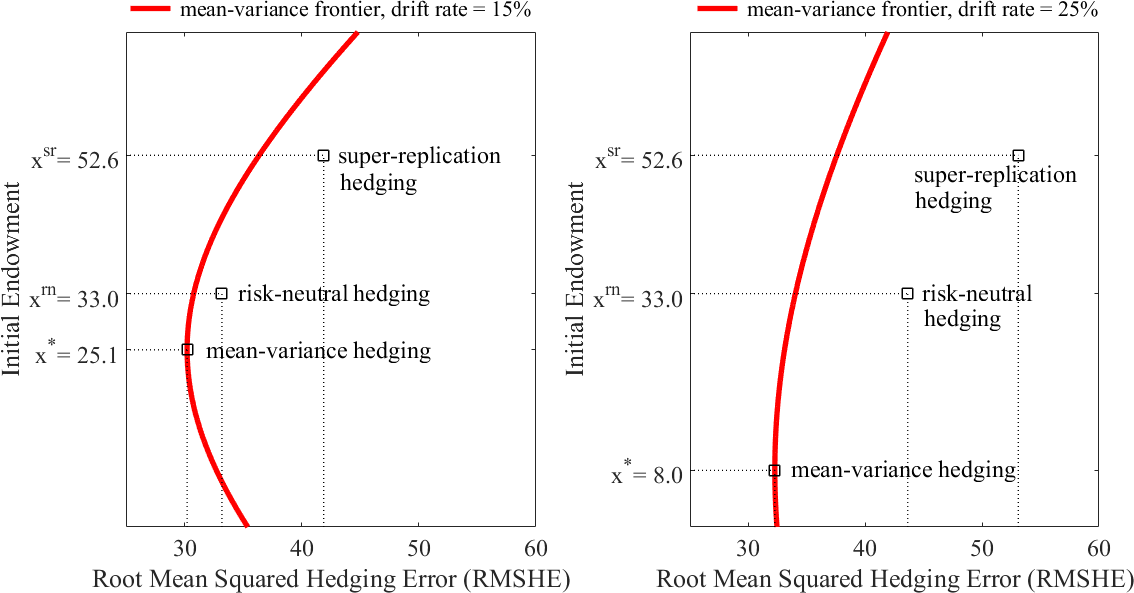}}
\caption{Mean-variance frontier, initial endowments and RMSHEs.
\label{fig:parabola}}
\end{center}
\end{figure}

We have numerically investigated the stochastic control
problems given by (\ref{P1}) and (\ref{P2}) by solving their
discrete time counterparts that arise if we approximate
the geometric Brownian motion $S$ by a binomial tree with
1000 time steps.
To this end, we have used the same parametrisation as in
Section~5 of Carr and Linetsky~\cite{CL}.
In particular, we have considered an ESO granted at the
money ($S_0 = K = 100$), with a ten year maturity ($T=10$)
and with payoff function $F(s) = \max(s-K, 0)$.
Contrary to Carr and Linetsky~\cite{CL}, who consider
immediate vesting, we have assumed a vesting period
of three years ($\Tv = 3$).
The intensity function $\ell$ is given by
\be
\ell (T-\tau, s) \equiv \lambda (\tau,s) = \lambda_f
+ \lambda_e (\ln s - \ln K)^+{\bf 1}_{\{ \tau \leq T-\Tv \}}
, \quad \text{for } \tau \in [0,T] \text{ and } s > 0 ,
\ee
for $\lambda_f = \lambda_e = 10\%$.
The constants $\lambda_f$ and $\lambda_e$
account for the ESO holder's job termination risk and
the fact that the ESO holder's desire to exercise
increases as the option's moneyness increases,
respectively.
We have assumed that the risk-free rate is
$5\%$ and the stock price volatility is $30\%$.
Furthermore, we have considered a drift rate of $15\%$
as well as  a drift rate of $25\%$.

\subsection{The mean-variance frontier}

We have solved numerically the recursive equations
associated with the discrete time approximation of the
problem given by (\ref{P2}).
For each $x$, we have thus computed the expected
squared hedging error at the random time of the ESO's
liquidation over all self-financing portfolio strategies
with initial endowment $x$.
The red parabolas in Figure~\ref{fig:parabola}, which we
call ``mean variance frontiers'', are plots of the square root
of this error, to which we refer as the ``root mean squared
hedging error'' (RMSHE), against the value of the initial
endowment $x$.
The ESO's mean-variance hedging value at time 0
is denoted by $x^\star$ and corresponds to the apex of
each parabola.

We have also used backward induction to compute the
ESO's risk-neutral value $x^{\mathrm{rn}}$, which has
been proposed by Carr and Linetsky~\cite{CL}, as well
as the ESO's super-replication value $x^{\mathrm{sr}}$
(see also (\ref{SRV}) and (\ref{RNV}) in the introduction).
In each of these two cases, we have computed the
corresponding RMSHEs using Monte Carlo simulation
along the lines described in the next subsection,
and we have located the associated points in
Figure~\ref{fig:parabola}.

As expected the ESO's risk-neutral and
super-replication values $x^{\mathrm{rn}}$ and
$x^{\mathrm{sr}}$ do not depend on the drift rate.
On the other hand, the ESO's mean-variance value
$x^\star$ is sensitive to the value of the market price
of risk.
Indeed, the two plots in Figure~\ref{fig:parabola}
illustrate the dramatic effect that the value of the drift
rate may have on the ESO's mean-variance hedging
value.

\subsection{Distribution of the hedging errors}

In the case of mean-variance valuation, we have considered
the portfolio strategy that starts with initial capital $x^\star$.
In the case of risk-neutral valuation, we have considered
the standard Black and Scholes Delta hedging strategy
with initial endowment $x^{\mathrm{rn}}$.
In the case of super-replication valuation, we have
considered the portfolio strategy that starts with
$x^{\mathrm{sr}}$ and hedges the American option that
yields the payoff $F(S_\tau)$ if exercised at time $\tau \in
[\Tv, T]$.
In each of the three cases, we have used Monte Carlo
simulation to compute the associated discounted to time
0 hedging errors, namely, the differences of the portfolios'
values and the ESO's payoff at the ESO's random
liquidation time.
We have derived empirical distributions of these hedging
errors using $50$ million samples (with this number of
samples, the simulated mean squared hedging error
of the mean-variance hedging strategy matches its theoretically
computed one up to the second decimal point).

\begin{figure}[!b]
\begin{center}
\scalebox{\figscaleB}{\includegraphics{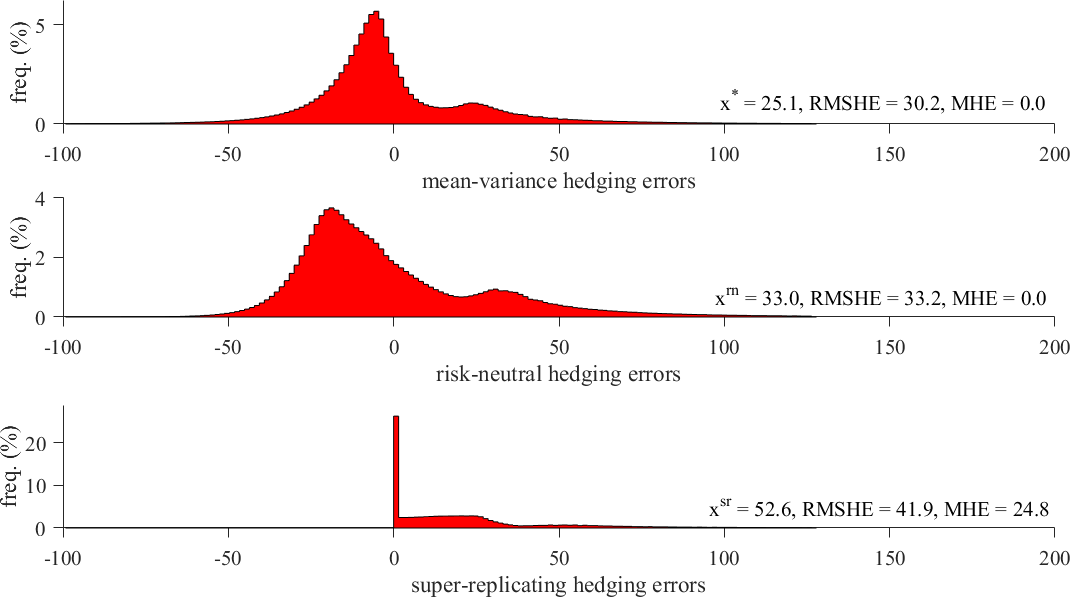}}
\caption{Histograms of hedging errors ($\mu = 0.15$)}.
\label{fig:histogram1}
\end{center}
\end{figure}

\begin{figure}[!t]
\begin{center}
\scalebox{\figscaleB}{\includegraphics{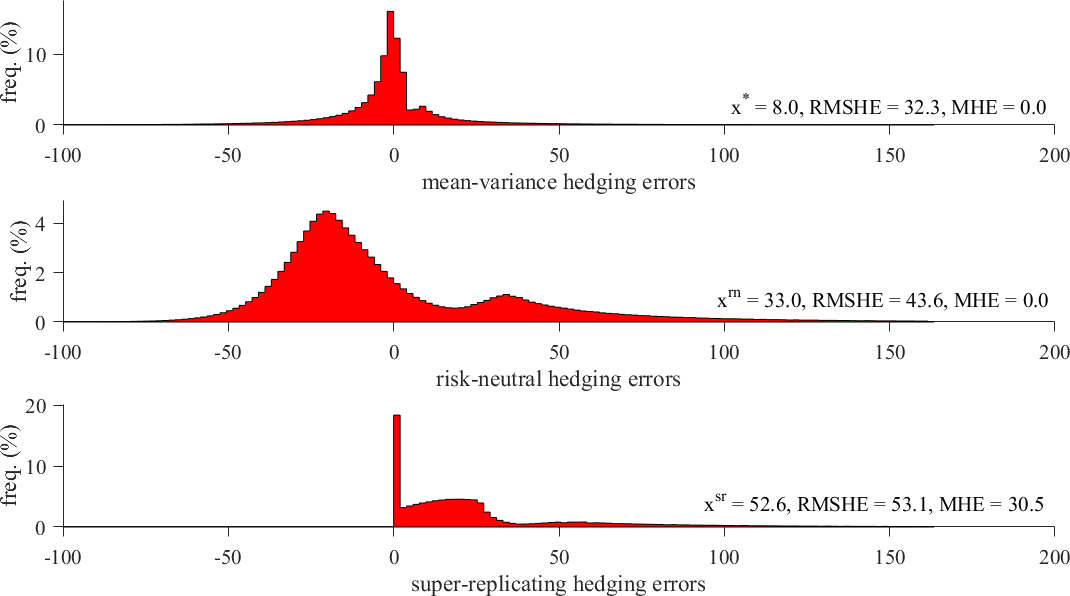}}
\caption{Histograms of hedging errors ($\mu = 0.25$).}
\label{fig:histogram2}
\end{center}
\end{figure}

We plot the empirical distributions of the hedging errors
for $\mu = 15\%$ and $\mu = 25\%$ in
Figures~\ref{fig:histogram1} and~\ref{fig:histogram2},
respectively.
In each of the six charts included in these figures,
we also mark the portfolios' initial endowments used,
namely, the values of $x^\star$, $x^{\mathrm{rn}}$ or
$x^{\mathrm{sr}}$, as well as the corresponding
mean hedging error (MHE) and root mean squared
hedging error (RMSHE).
Furthermore, we report selected percentiles of the hedging
errors  in Table~1.

We note that a negative value of the hedging error means
that the portfolio's value has not covered the ESO's
payoff.
As expected, the super-replication strategy never leads to
a negative hedging error.
The ``hump'' that appears in the frequency of positive
hedging errors is due to the fact that, if the ESO liquidation
occurs during the vesting period, then the ESO forfeits
without yielding a payoff.
Indeed, if the vesting period is changed from three years
to immediate vesting, then the bimodality of the hedging
error distribution disappears.

\begin{center}
\small
\begin{tabular}{cddddddddd}

\mc{endowment}  &  \mc{\ MHE} & \mc{RMSHE} & 
\mc{1\%}&\mc{5\%}&\mc{10\%}&\mc{50\%}&\mc{90\%}
&\mc{95\%}&\mc{99\%}\\\hline\hline

\multicolumn{10}{l}{drift rate set to $15\%$}\\
$x^{\star\ }       = 25.1$ & 0.0 & 30.2 & -63.4 & -34.4 & -24.6 &
-5.4 & 32.2 & 50.4 & 108.0 \\
$x^{\text{rn}} = 33.0$ & 0.0 & 33.2 & -47.1 & -34.5 & -29.0 &
-9.3 & 41.2 & 60.2 & 117.4 \\
$x^{\text{sr}} = 52.6$ & 24.8 & 41.9 & 0.0 & 0.0 & 0.0 &
15.4 & 64.4 & 88.9 & 157.0\\\hline

\multicolumn{10}{l}{drift rate set to $25\%$}\\
$x^{\star\ }       =  8.0$ & 0.0 & 32.3 & -79.7 & -34.2 & -20.0
& -1.4 & 18.2 & 40.4 & 119.1 \\
$x^{\text{rn}} = 33.0$ & 0.0 & 43.6 & -58.2 & -42.4 &
-35.4 & -13.7& 52.2 & 80.5 & 161.7 \\
$x^{\text{sr}} = 52.6$ & 30.5& 53.1 & 0.0 & 0.0 & 0.0 &
17.4& 78.8 & 113.9 & 208.6 		

\end{tabular}
\bigskip
\normalsize

{\bf Table 1:} Mean hedging errors (MHE), root mean
squared hedging errors (RMSHE) \\
\mbox{}\hspace{20mm} and $1\%$, $5\%$,
$10\%$, $50\%$, $90\%$, $95\%$, $99\%$ percentiles
of the hedging errors.
\end{center}

\subsection{Convergence for long time horizons}

\begin{figure}[!t]
\begin{center}
\scalebox{\figscaleB}{\includegraphics{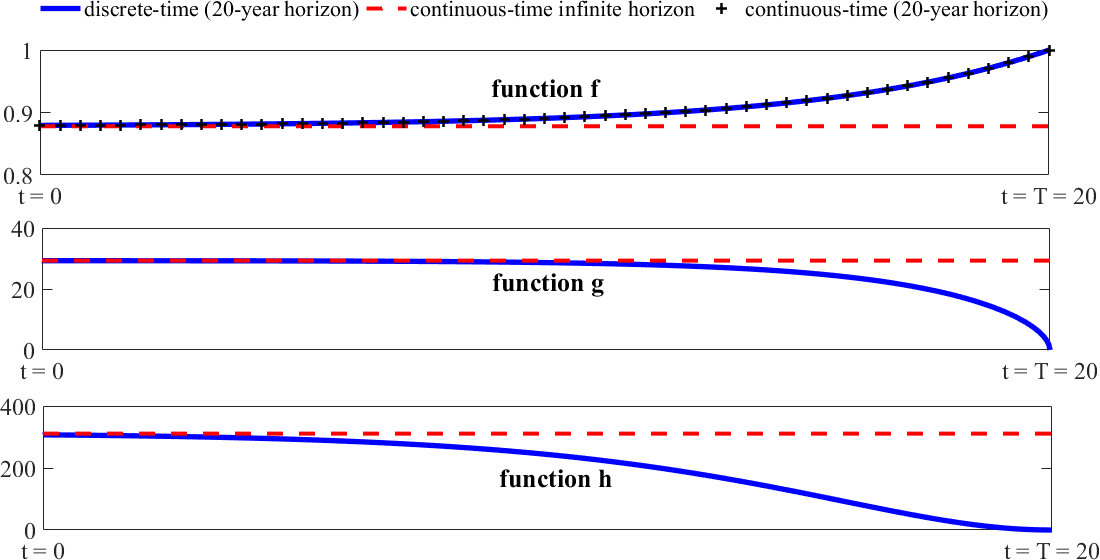}}
\caption{Illustration of convergence for long time
horizons.\label{fig:infiniteT}}
\end{center}
\end{figure}

To illustrate the convergence of the mean-variance
valuation scheme as time to horizon becomes large,
we have considered the ESO described at the beginning
of the section but with a twenty year maturity ($T=20$)
and with immediate vesting.
We have also assumed that $\lambda_f = 20\%$,
$\lambda_e = 0$,  $\mu = 10\%$, $\sigma = 30\%$
and $r=5\%$.
Such choices put us in the context of (\ref{par-ineq-oo-1})
in Remark~\ref{rem:infinite}.
In Figure~\ref{fig:infiniteT}, we plot the functions
$f$, $g$ and $h$ as computed using the binomial tree
model.
In the first chart of the figure, we also plot the
function $f$ arising in the context of the continuous time
model, which is given by (\ref{f-lam-const}).
We also plot the level given by the functions $f_\infty$,
$g_\infty$ and $h_\infty$ evaluated at the initial stock price
$S_0 = 100$ using the closed form formulas derived in
Remark~\ref{rem:infinite}.

%==========================================
\section*{Acknowledgement}

\noindent
We are grateful to Monique Jeanblanc for several
helpful discussions and suggestions.
The first author would like to thank the Norwegian School
of Economics, the Norske Bank fond til {\o}konomisk forskning, and Professor Wilhelm Keilhaus's minnefond for supporting his research stay at the LSE.

\end{document}